# Guiding Large Language Models to Generate Computer-Parsable Content

**Jiaye Wang**
hk-shao@outlook.com
School of Software, South China Normal University
Platform and Content Group, Tencent Inc.

March 26, 2024[1]

## Abstract

Large language models (LLMs) have demonstrated remarkable capabilities in learning patterns from massive text corpora, including word relationships, sentence structures, and even complex semantic and pragmatic information. However, it remains challenging to induce pre-trained language models to generate structured content that strictly follows specific conventions.

We propose a scheme for guiding LLMs to generate highly usable content for computers without the need for fine-tuning and additional neural network inference, by introducing coroutine-based content generation constraints through a pre-agreed context-free grammar (CFG), which guides the autoregressive model Transformer to sample the correct tokens during its decoding phase to form a program-compliant form in the decoding phase of the autoregressive model Transformer to form a formal language that conforms to the program conventions. This will effectively improve the stability and consistency of LLMs in generating target data structures, types or instructions, and reduce the difficulty of application development and integration.

We first conducted the matching bracket pairs experiment to verify that the error rate of models like GPT-2 and Gemma reaches 95% when the generated DSLs exceed lengths of 36 and 282 characters, respectively. This finding highlights the performance limitations of some current LLMs in generating specific DSLs. We introduce YieldLang, a coroutine-based framework for DSL generation. We then conducted experiments using LLMs on multiple datasets encompassing tasks such as JSON, Mermaid flowchart, and function call expression generation. These experiments demonstrate that our approach improves accuracy by a factor of 1.09 to 11.6 compared to benchmarks. In the best case, it reduces the number of samples required by LLMs for JSON generation to approximately 16.5% of the benchmark level. This significantly improves the usability of LLM-generated content for computer programs.

[1]First published at: https://chinaxiv.org/abs/202403.00340

# 1 Introduction

## 1.1 Background and Significance

LLMs have allowed us to witness computers possessing a certain degree of ability to understand and generate natural language, along with a certain generalization capability across different domains. This human-machine dialogue capability can help humans solve various problems. However, computer programs written by developers do not have the same level of robustness as humans. Under the premise of providing reasonable input prompts, there are two pathways to enable pre-trained LLMs to more accurately guide the completion of a series of automated tasks or be incorporated as middleware into more types of production processes in computer programs.

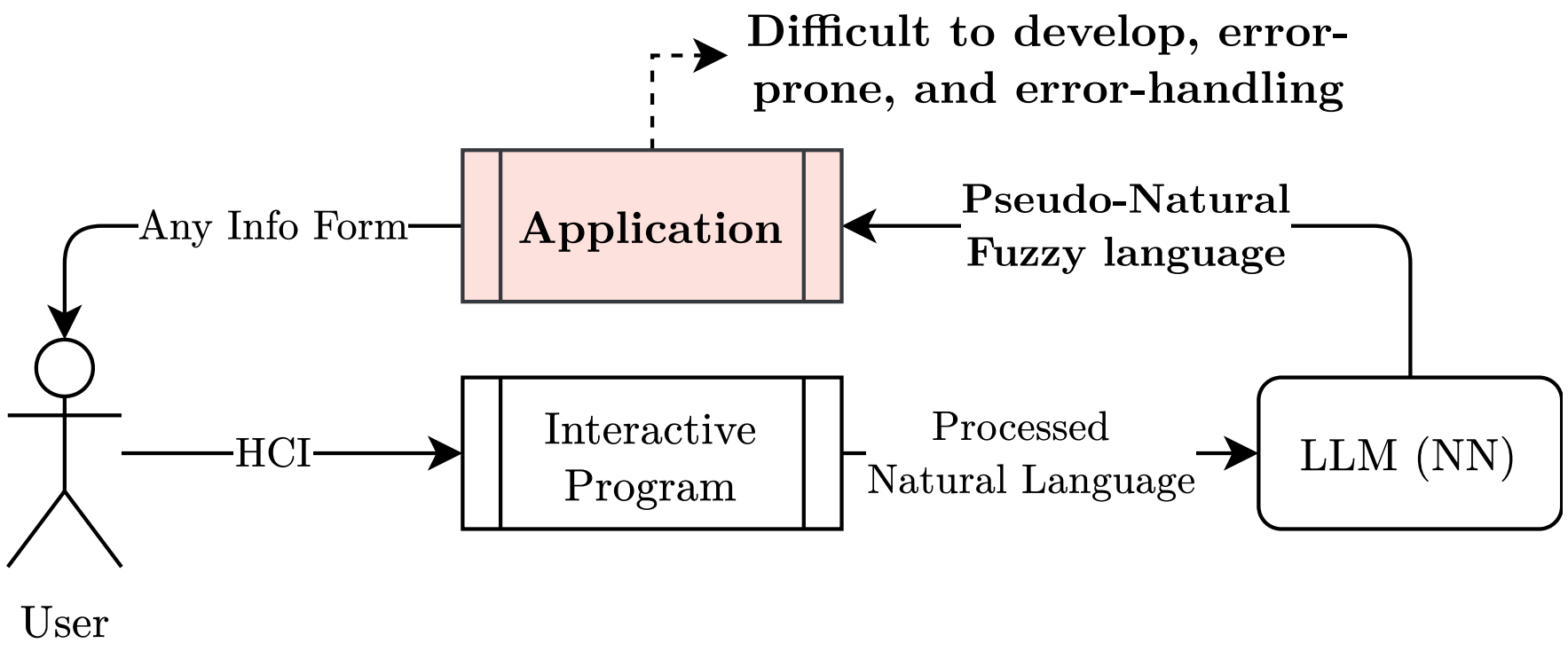

Figure 1: Poorly structured content is not easily processed by applications

There are two common approaches, which are developing more robust applications and aligning language models. However, both approaches may have certain issues, specifically:

1. Language models such as GPT possess language understanding and generation capabilities that enable the development of applications closer to artificial general intelligence. However, if developers need to build applications based on LLMs as shown in Figure 1, they often encounter some pain points. The output of LLMs depends on probability, and users or computer programs cannot ensure with 100% certainty that the content generated by LLMs conforms to the expected structure. It is more difficult and costly for developers to develop more robust programs to handle the model's output. Unexpected outputs often lead to more information loss, computational resource waste, and even failure of the entire task chain.
2. Another approach is to align the language model, making it more adapted to our applications for specific tasks. This can also improve the usability of the generated content and reduce the waste of computational resources. However, aligning language models incurs an "alignment tax" (LLMs gained broad capabilities during pretraining, but reinforcement learning under human feedback to align models leads to forgetting) (Lin et al. 2024), and models newly aligned for a class of tasks require new deployments, resulting in significantly increased computational resource costs.

## 1.2 Related Work

Language models are not only capable of direct human-machine interaction, but they should also possess the ability to accurately utilize given tools, which is formally equivalent to enabling language models to understand instructions and then use the provided application programming interfaces (APIs). This essentially requires the output of language models to be parsable by computers into domain-specific languages conforming to the expected data structures. Relevant engineering and research efforts have been continuously emerging in both academia and industry in recent years.

### 1.2.1 Research on Controlling Language Models in Academia

In February 2018, OpenAI released the open-source pre-trained language model GPT-1 through a paper titled "Improving Language Understanding by Generative Pre-Training" (Radford et al. 2018), followed by GPT-2 (Radford et al. 2019) in 2019 and GPT-3 in 2020. Initially, these three versions were only well-known in academia until OpenAI launched the commercial product ChatGPT on November 30, 2022. In the following short span of just 5 days, it garnered 1 million registered users, and two months later, its monthly active users reached 100 million, becoming a hot topic in the business and tech world. OpenAI opened a web interface, allowing developers to use OpenAI's language models to develop "more incredible" applications. More researchers have participated in the application research of LLMs, finding that through a prompting technique, language models can "reason" through "thought chains" and "act" by utilizing predefined tool sets (such as searching the internet). Experiments have proven that this combination greatly improves the quality of output text and endows LLMs with the ability to correctly solve problems (Yao et al. 2023).

Noting the demand from numerous developers for more flexible control over the GPT, OpenAI officially released GPT-4 Turbo at the OpenAI DevDay on November 6, 2023. In the Opening Keynote, OpenAI's CEO Sam Altman highlighted six major updates, with a focus on the "More Control" section introducing the JSON mode. This feature allows GPT to directly generate JSON (JavaScript Object Notation) text. Unlike ambiguous natural language, JSON belongs to a context-free language widely adopted in industry and easily parsed by programs into definite data structures, thereby better participating in program logic flows.

Researchers at ElementAI found that adopting Parsing Incrementally for Constrained Auto-Regressive Decoding (PICARD) can effectively enable language models to generate formal languages, especially valid SQL query statements (Scholak, Schucher, and Bahdanau 2021). This approach of rejecting illegal words during autoregressive decoding and outputting only legal DSL prefixes may be the methodology behind OpenAI's implementation of the JSON mode (Lages 2024). Apart from the JSON mode, function calling also allows GPT to generate structured outputs. By providing the function name, description, required parameters, and corresponding types in advance, GPT generates function call information in JSON format, which is still a form of "JSON mode" but endows GPT with the ability to use tools.

In late May 2023, researchers from Google and MIT proposed Grammar Prompting, which first prompts LLMs to generate BNF grammar rules based on context, and then uses constrained decoding to generate text following those grammar rules. Through experiments on semantic parsing DSLs (SMCalFlow, GeoQuery, Overnight), action DSLs (PDDL planning), and molecular generation (SMILES), the researchers believe Grammar Prompting can improve upon standard prompting baselines, and this approach holds promise for better assisting LLMs in using tools (Wang et al. 2024).

### 1.2.2 Demand for Generating Structured Strings in Industry

In industry, a typical demand for LLMs to generate structured text is to produce JSON strings. Some developers have realized that LLMs cannot reliably generate valid JSON strings, but fortunately, the errors made by LLMs tend to be relatively typical. As a result, in most cases, a parsing program can repair the errors produced by LLMs without compromising the meaning (Baccianella, Powley, and Terry 2024). This approach involves first allowing LLMs to generate content and then attempting to repair it. However, the program may not be able to adapt to various large models and repair strings in all situations.

On November 11, 2022, developers from Microsoft committed the first line of code for guidance, a Python software package for controlling LLMs, to the code hosting platform GitHub. Through continuous updates and iterations, guidance provides ways to control language models from interfaces such as Transformers, LlamaCpp, VertexAI, and OpenAI using regular expressions, context-free grammars, and interleave control, allowing developers more flexible control over language models (Lundberg et al. 2024).

The Transformers library, open-sourced by Huggingface on GitHub, provides numerous pre-trained language models and source code implementations (Wolf et al. 2024). On November 17, 2023, developer Saibo Geng from EPFL submitted a PR (Pull Request) #27557 to the Transformers repository, providing a draft implementation of "Grammar-Constrained Decoding" using the EBNF interface. Saibo Geng argued that grammar-constrained language models significantly outperform unconstrained language models and even outperform fine-tuned models for specific tasks. Grammar constraints have great potential for leveraging off-the-shelf language models for a wide range of structured NLP tasks, especially when training data is scarce or fine-tuning costs are high (Geng et al. 2024).

The more commonly used vLLM library provides a fast and flexible open-source LLM inference service implementation (Kwon et al. 2024). On December 14, 2023, developer Andrew Lapp[2] committed PR #2105 to the vllm repository, implementing an incremental LALR algorithm for a CFG or regular expression parser to determine the set of "legal-next-tokens" during the language model generation process. This PR depends on the open-source grammar parser Lark (Shinan et al. 2024) and implements the `vllm.grammar.GrammarLogitsProcessor` post-processor with the EBNF interface to constrain LLMs to generate content that strictly conforms to the specified grammar conventions.

### 1.2.3 Limitations of Existing Research and Technical Solutions

The problems with existing research and technical solutions can be summarized in three main points: high training and inference costs, lack of generality and usability, and poor robustness and time complexity.

1. The training and inference costs of LLMs are high. The process of training or fine-tuning models for different DSL generation requirements is cumbersome and costly, and there is still no guarantee that the generated content will achieve 100% usability. If the generated content does not conform to the expected grammar and fails the syntax check, it can only be regenerated or terminated as a failure. Industrial solutions for JSON repair cannot handle all cases and are not applicable to other DSLs, imposing a significant cognitive burden on developers who need to write such "repair" programs.

[2]https://github.com/lapp0

2. Parser generators like Antlr4, Lark, and Tree-sitter (Brunsfeld et al. 2024) are designed to generate high-performance parsing (or incremental parsing) code that converts a string conforming to a grammar into an Abstract Syntax Tree (AST). Their design goals do not include enhancing robustness or aiding in the automatic generation of DSLs. When the input text violates the grammar, these programs typically fail and terminate, providing error messages but lacking precise information about the location and cause of the error, which is insufficient for guiding DSL generation. These parsers are also unable to recover from a partial parse and guide DSL generation.

3. The CFG-related proposals in the Transformers and vLLM repositories, as well as relevant research from journals and conferences, do not adopt an "asynchronous" + "instructional" approach. Even with incremental parsing, in the worst case, existing techniques require verifying whether the currently generated prefix belongs to the DSL or its prefix every time the autoregressive language model generates a token. Some optimizations, such as function call caching and incremental parsing, have been proposed, but the time complexity may still be suboptimal. If generating a DSL of length $n$ and verifying its membership in the DSL has a time complexity of $O(f(n))$, then the complexity of generating the DSL would be $\sum_{i=1}^{n} O(f(i))$. Depending on the DSL or parser, $O(f(n))$ could be $O(n), O(n^2)$, or $O(n^3)$, among other possibilities.

## 1.3 Research Content

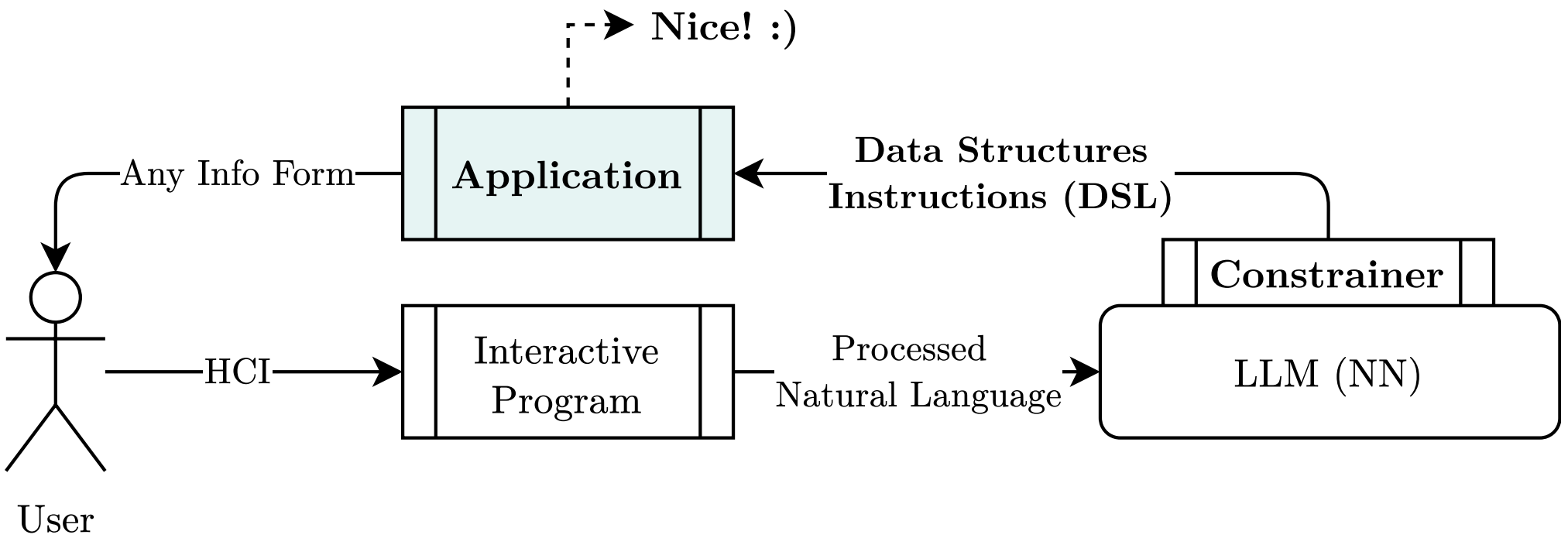

Figure 2: Constraints to generate structured data or instructions

Firstly, developers need to integrate language models into application implementations with lower cognitive burden. In the model shown in Figure 2, users neither directly interact with LLMs (which can reduce the insecurity of LLMs being "brainwashed" by attackers) nor directly accept outputs from LLMs. Instead, the outputs are processed by the application and transformed into other forms of human-computer interaction (e.g., responding to users through charts, animations, and sounds).

Secondly, to address the limitations of existing research and technical solutions, the proposed guiding module consists of two sub-modules: Module one is an asynchronous and coroutine-based DSL parsing and generation module, termed the "Asynchronous DSL Parsing Language Generation Module"; Module two is a decoding module that invokes the language model for DSL sampling based on Module one, termed the "Language Model Guiding Module". Through these two modules, LLMs are guided to generate the expected DSL at appropriate positions, allowing developers to more easily leverage the capabilities of LLMs to develop more robust applications. The overall system workflow is roughly illustrated in Figure 3.

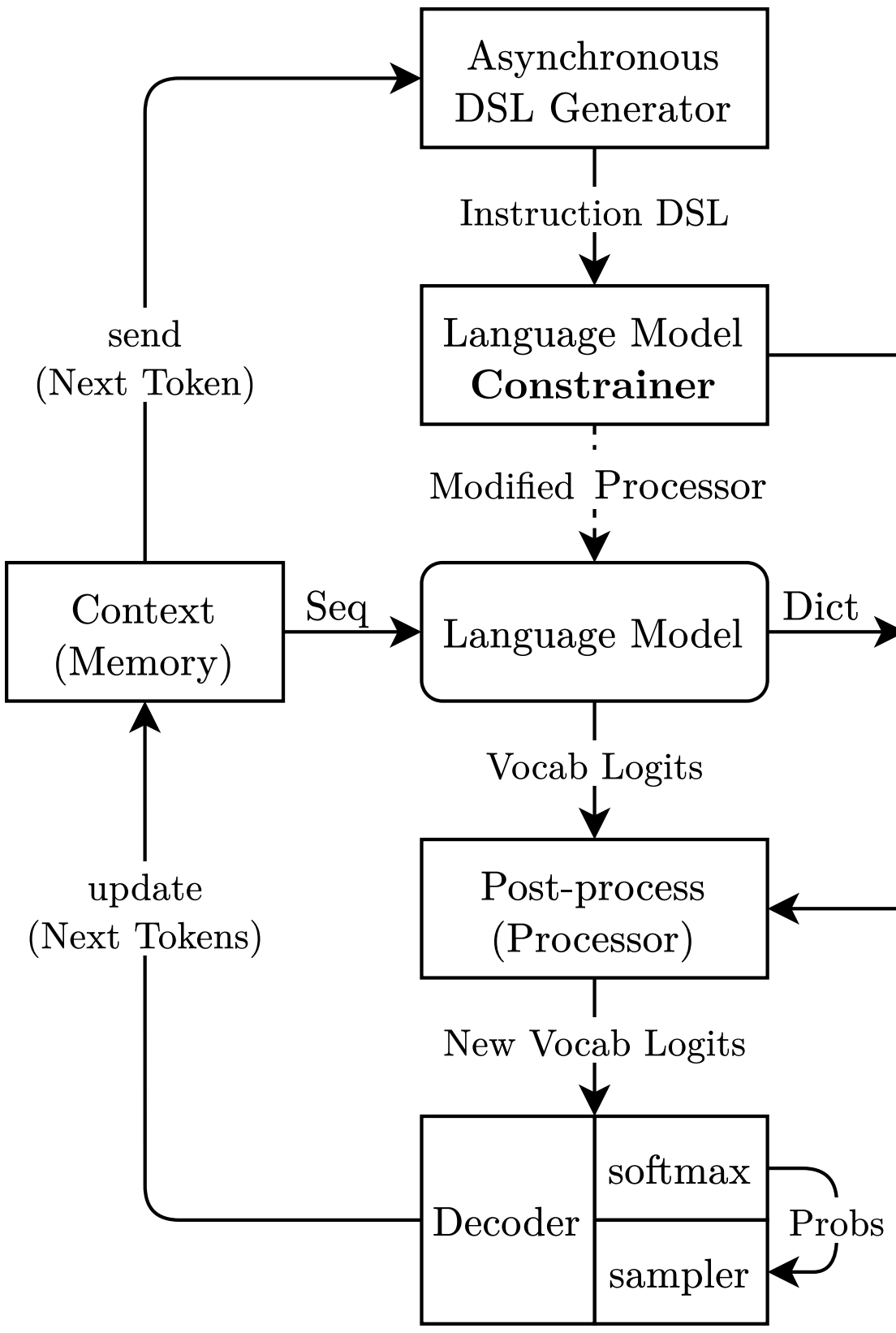

Figure 3: A system for guiding language models

1. Based on the framework and the set of combinators, decorators, and generators provided by this study, developers use programming concepts such as Python coroutines, classes, functions, and recursion to write generator classes for context-free grammars (CFGs) in Backus-Naur Form (BNF). This generator can produce prompts to guide the generation of the language corresponding to the given CFG.

2. An apparatus runs the developer-provided generator class to produce prompting instructions and drive Apparatus 2, which in turn drives an autoregressive language model to sample the next legal token from the language prompts. Specifically, for Transformer-based models, Apparatus 2 disables all invalid tokens (token IDs corresponding to the language model's vocabulary) in the "post-processing" stage before decoding. Then, the language model's decoding module produces the next token (string type), selecting the legal token prefix that can form the next DSL prefix.

3. The generated next token prefix is incorporated into the context of the autoregressive language model, and this token prefix is also input to Apparatus 1, which then produces the next prompting instruction. This process continues until Apparatus 1 produces a termination instruction (e.g., EOF, EOS), at which point the DSL generation is complete.

## 1.4 Chapter Structure

This paper will mainly introduce the structure and generation methods of domain-specific languages (DSLs) commonly used in industry, how to design an asynchronous DSL generation and parsing apparatus implemented with coroutines, and how to build a constrained decoding apparatus for autoregressive language models (especially language models with the Transformer architecture) based on this apparatus, so that developers can more easily leverage the capabilities of LLMs to develop applications with higher robustness on the basis of this apparatus. Specifically, the arrangement of each chapter is as follows.

Chapter 1: Introduction. The main content is the background of this research, the difficulties of incorporating LLMs into applications, research on controlling language models in academia, the demand for generating structured strings in industry, the limitations of existing research and technical solutions, the research content of the topic, and the structure of this paper.

Chapter 2: Developers Need Language Models to Generate DSLs for Application Development. It mainly introduces the forms of DSLs commonly used in industry, academia and other fields. Computers rely on DSLs to implement specific functions, while developers need to rely on DSLs and utilize the general capabilities of language models to develop applications.

Chapter 3: Performance Deficiencies of Large Language Models in DSL Generation Tasks. It mainly introduces the poor performance of existing LLMs in generating DSLs according to instructions, and their inability to consistently generate text conforming to the expected structure. This chapter also analyzes the reasons for these performance deficiencies from the decoding strategies of existing language models.

Chapter 4: Basic Rules and Paradigms for Generating DSLs, and Implementation of Generators. It mainly introduces the generation rules of formal languages, treating DSLs as a form of formal language, especially the most commonly used context-free languages. This chapter introduces a DSL generation apparatus implemented with recursive descent and coroutines, which can also be used for DSL parsing and verification. Through random sampling or other sampling methods, this apparatus can generate content that can be parsed by computers.

Chapter 5: Generating Expected DSLs by Developers through Language Model Sampling. It mainly introduces a constrained decoding method that uses existing Transformer-based language models as samplers to select DSL generation paths in the generator. This method can also be applied to other autoregressive language models.

Chapter 6: Advantages of This Research in Language Model DSL Generation. It mainly introduces the experimental results of this research, conducting experiments on multiple task datasets including DSL generation tasks such as JSON, Mermaid flowcharts, and function call expressions, as well as testing the impact of the DSL constraint apparatus on the number of samples used by language models.

Chapter 7: Summary and Outlook. It summarizes the work done in this research, looks forward to the applications of the method proposed in this research, and points out the limitations of this research and possible future research directions.

## 1.5 Summary of This Chapter

In this chapter, we summarize the challenges developers may face when using LLMs to develop specific applications: it is difficult to develop sufficiently robust computer programs to parse the output content of language models into the data structures expected by developers and programs. There are currently three approaches in academia and industry to address this issue, including "post-processing," "model alignment," and "constrained decoding." However, existing methods have certain limitations. This research aims to improve the existing implementation of "constrained decoding" by implementing an asynchronous DSL parsing language generation module and a language model guidance module. Developers can more easily leverage language models to generate DSL to develop more robust applications.

# 2 Developers Need Language Models to Generate DSLs

Domain-Specific Languages (DSLs) have been widely applied in programming practice. Developers rely on DSLs for data exchange, configuration file writing, and business logic implementation. Parsing and generating DSLs is a standard functionality provided by standard libraries, packages, or application development frameworks across various programming languages. With the advancement of LLMs, people have discovered that LLMs have broad applications in semantic understanding and generation. Developers hope to leverage the general capabilities of LLMs to develop new applications, and DSLs are an important avenue for LLMs to interact with existing applications. Therefore, developers need language models to generate DSLs for application development.

## 2.1 Domain-Specific Languages

A Domain-Specific Language (DSL) is a computer language designed for a specific domain application, also known as an application-oriented language (Sammet 1969) or special-purpose language (Wexelblat 1981). Unlike General-Purpose Languages (GPLs), DSLs focus on the expressive power of a specific domain and provide a more concise and comprehensible syntax and semantics.

For the sake of convenience, the DSLs referred to in this paper emphasize strings that can be parsed by computer programs into specific data structures, or called formal languages, without distinguishing between DSLs and GPLs. Languages used for data exchange such as JSON, XML, YAML, query languages such as SQL, XPath, XQuery, markup languages such as HTML, LaTeX, Markdown, as well as various programming languages, are all considered DSLs.

## 2.2 Industrially Common DSLs

In addition to using general-purpose programming languages (such as C, Java, Go, and Rust), software developers often need to use various DSLs during the development process. These DSLs may involve domains including markup languages, information exchange languages, data formats, application interfaces, software configurations, typesetting, and more. Table 1 lists some common industrial DSL names, MIME types, and brief introductions.

| Names | MIME Types | Brief Introductions |
|---|---|---|
| JSON | `application/json` | (Bray 2014) JSON defines a lightweight set of rules for portably representing structured data. JSON's grammar is context-free, and it is one of the most commonly used data exchange formats on the Web, widely applied in frontend-backend data exchange, configuration file writing (e.g., the `package.json` file in NodeJS), and more. |

| Names | MIME Types | Brief Introductions |
|---|---|---|
| XML | `application/xml` | (St.Laurent, Makoto, and Kohn 2001) The Extensible Markup Language (XML) is a markup language that can be used for storing, transmitting, and reconstructing loosely structured data. XML's grammar is context-free, and it is widely used in frontend-backend data exchange, configuration file writing (e.g., the `pom.xml` file for Maven packages), and more. |
| HTML | `text/html` | (Masinter and Connolly 2000) The HyperText Markup Language (HTML) is a markup language used for creating web pages. HTML's grammar is context-free, and it is widely applied in typesetting and frontend application development. |
| CSV | `text/csv` | (Shafranovich 2005) Comma-Separated Values (CSV) is a common, simple file format used to store tabular data of numbers or text in plain text. It serves as an information exchange format widely used by users, businesses, and the scientific community. |
| Mermaid[3] | `text/vnd.mermaid`[4] | (Sveidqvist and Mermaid 2014) A JavaScript-based diagramming and charting tool inspired by Markdown's syntax and rendering, used to create and modify various diagrams including flowcharts, sequence diagrams, Gantt charts, class diagrams, Git graphs, entity relationship diagrams, etc. |

Table 1: Examples of industrial DSLs

The MIME in Table 1 refers to Multipurpose Internet Mail Extensions, a standard for indicating the content type of Internet transmissions, which is also a DSL defined by BNF (Freed and Borenstein 1996), initially defined in the Request for Comments (RFCs) .

## 2.3 DSLs Used by Academia

In academia, the most common requirement is for a DSL capable of accurately and strictly representing mathematical or other disciplinary formulas, which should be easy for users to input. It should be parsable by corresponding computer programs into specific data structures and then rendered into vector graphics or bitmaps for screen display or printing via typesetting and rendering algorithms capable of handling complex mathematical symbols. The most classic DSL for mathematical typesetting is TeX, followed by MathML (a markup language for describing mathematical notation by capturing both its structure and content, which is compatible with the HTML5 standard) (Ion et al. 1998) and the emerging Typst programming language for typesetting (Mädje 2022).

In the field of physics, Q# is a quantum algorithm toolkit developed by Microsoft, later formalized by Kartik Singhal et al. from the University of Chicago, who proposed λ-Q#, with a more rigorous mathematical definition and type system (Singhal et al. 2023), providing a DSL with a well-defined syntax for describing quantum algorithms.

[3] https://mermaid.js.org/intro/

[4] https://www.iana.org/assignments/media-types/application/vnd.mermaid

Some researchers have created a specialized markup method, SMILES, to represent chemical molecular formulas (Weininger 1988). This representation uses ASCII strings to explicitly specify molecular structures, and some researchers have used deep neural networks to train probabilistic models for generating SMILES strings to produce chemical structures (O'Boyle and Dalke 2018). In 2007, the open-source chemistry community developed OpenSMILES[5], an open specification for which the language was defined using a BNF grammar (Vandermeersch et al. 2021).

In the field of biology, researchers employ deep learning, natural language processing, and other techniques to study proteins, as they believe that the sequences and structural domains of common proteins are analogous to words, phrases, and sentences in human language (Ofer, Brandes, and Linial 2021), and DSLs can thus be used as tools for protein research.

## 2.4 DSLs in Other Domains

DSLs are also used in other fields, such as business and art. COBOL (Common Business Oriented Language), a specialized programming language for business, was one of the earliest high-level programming languages and the first standardized computer language. DSLs like ChucK (Wang, Cook, and Salazar 2015) for real-time sound synthesis or music creation, Csound (Lazzarini et al. 2016) (an XML-based sound computing system), and Processing (a programming language designed for electronic arts and visual interaction) (Bohnacker et al. 2012) have found widespread applications in their respective domains.

## 2.5 Language Models for Generating DSLs

Language models like GPT have demonstrated their general capabilities in various domains, including machine translation, text summarization, information extraction, question answering, code generation, and more. More specifically, we hope that language models can perform tasks such as recognition, classification, routing, reasoning, and decision-making in specific environments. When language models are involved in concrete computer applications or research, generating specific DSLs is necessary for computer programs to recognize them. If LLMs can generate DSLs by following instructions, their usability and versatility will be greatly enhanced. This will be one of the paths towards enabling artificial intelligence to learn how to use tools and progress towards Artificial General Intelligence.

## 2.6 Summary of This Chapter

In domains where computers intersect with industries, academia, business, and art, DSLs have gained widespread adoption, and a number of researchers have employed language models for academic investigations. Language models like GPT possess remarkable versatility, while developers rely on DSLs. If language models could accurately follow instructions and generate DSLs as desired by developers, they would become more practical and have an exceptionally broad range of applications.

[5] http://opensmiles.org/opensmiles.html

# 3 Performance Deficiencies of LLMs in DSL Generation

Although domain-specific languages (DSLs) find wide applications across various domains, generating valid DSLs using deep learning models poses a challenging task. For instance, the authors of DeepSMILES mentioned in their 2018 paper that employing a character-level recurrent neural network (RNN) model to generate SMILES strings might lead to generating numerous invalid strings, primarily due to the model's inability to produce matching bracket pairs (O'Boyle and Dalke 2018; Wolfram 2023). These invalid strings cannot be parsed into legitimate chemical structures by chemical software.

Many DSLs specify matching bracket pairs, quotation pairs, or other language structures at the grammar level, upon which corresponding software relies for DSL parsing. If LLMs are to be used for DSL generation, it is crucial to first verify whether existing LLMs can generate the prescribed DSLs according to instructions and assess their performance.

## 3.1 Common DSL Syntactic Rules

The most common syntax involves matching bracket pairs, where the fundamental rule dictates that each left bracket is matched with a corresponding right bracket. Such syntactic structures are prevalent in numerous programming languages or DSLs, including JavaScript, Python, Lisp, and JSON, among others. If the alphabet $\Sigma$ comprises only two characters, `'('` and `')'`, the simplest bracket pair is `'()'`. Given a string $s$ in $\Sigma^*$, a commonly used algorithm to verify if the string belongs to valid bracket pairs is to use a stack for validation. However, for cases with only two characters, it can be simplified to the function Legal($s$), as shown in Equation 1, to validate that $s$ is a valid string prefix.

$$\text{Legal}(s) = \bigwedge_{k=1}^{|s|} \left[ \left( \sum_{i=1}^{k} \begin{cases} +1 \text{ if } s[i] = \text{'('} \\ -1 \text{ if } s[i] = \text{')'} \end{cases} \right) \geq 0 \right] \tag{1}$$

When the alphabet $\Sigma$ includes multiple pairs of brackets, for example, $\Sigma = \mathbb{L} \cup \mathbb{R}$; $\mathbb{L} = \{$`'('`, `'['`, `'{'`$\}$; $\mathbb{R} = \{$`')'`, `']'`, `'}'`$\}$, if all valid strings $s \in \Sigma^*$ form the set $L(G)$, then Figure 4 shown below can verify whether a string $s$ belongs to the language $L(G)$, where $G$ is the grammar and $\varepsilon$ is the empty string. The function $f$ maps left brackets to right brackets.

*Legal-Bracket-Pairs* ($s \in \Sigma^*$, $f$: $\mathbb{L} \to \mathbb{R}$):

- $\sigma \leftarrow$ **empty stack** ▷ Initialize empty stack $\sigma$
- **for** $c$ **in** $s$**:** ▷ Traverse $\forall(i, c) \in s$ in order
  - **if** $c$ **in** $m$**:** ▷ The character $c$ is the left bracket
    - **push** $c$ **onto** $\sigma$ ▷ Push character $c$ onto stack $\sigma$
  - **else:** ▷ Character $c$ is other character
    - $l \leftarrow$ (**pop** $\sigma$) **if** $|\sigma| > 0$ **else** $\varepsilon$
    - **if** $f(l) \neq c$ **:** ▷ Character $c$ cannot match $l \in \mathbb{L}$
      - **return False** ▷ String $s$ is not a legal string
- **return** $|\sigma| \ngtr 0$ ▷ $\forall(i, l \in \mathbb{L}) \in s$ is legal if it is matched

Figure 4: Algorithm for checking if a pair of brackets matches

In addition to matching bracket pairs, escaping strings is also a common syntactic rule in various DSLs. When a string is enclosed within a pair of quotes, the characters within the string cannot directly contain quotes; they need to be escaped instead. For example, to represent a string containing a `"` character, it needs to be written as `"\""` instead of `"""`, otherwise, the syntax parser of the application will prematurely terminate the parsing of the string after processing the second `"` character, leading to information loss.

## 3.2 Experimental Approach to Validate DSL Generation Performance

Since matching bracket pairs and string escaping are widely present syntactic rules in DSLs, validating the DSL generation performance of language models can start from these two aspects. If language models cannot even adhere to these most basic syntactic rules, then the performance in generating more specific DSLs will obviously not be satisfactory.

As GPT and other common language models are autoregressive models, the generated text is produced token by token. For these language models, the prerequisite for generating a legal bracket pair $L(G)$ is to generate a prefix of a legal bracket pair string. Assuming the language model predicts only tokens $l =$ `'('` and $r =$ `')'`, according to Equation 1 and the rules of legal bracket pairs, it is known that $\mathbb{I} = \{srx \mid \forall s \in L(G), x \in \Sigma^*\}$ is the set of illegal bracket pairs.

| Input | ( | ) |
|---|---|---|
| (((()))) $x \rightarrow$ | 97.8% | 2.2% |
| ((((())))) $x \rightarrow$ | 99.9% | 0.1% |
| (((((())))))) $x \rightarrow$ | 97.4% | 0.6% |
| ((((((())))))))) $x \rightarrow$ | 97.9% | 2.1% |

Figure 5: Probability prediction of the next bracket by the GPT-2

The basic idea of the experiment is to construct the string $s = l^n r^n \in L(G)$, input the prompt word of Table 2 to let the model predict the probability of `'('` or `')'`, and The probability of $r =$ `')'` is exactly the error rate, as Figure 5 indicates.

| Chinese Prompt | English Prompt |
|---|---|
| 给定一个仅包含 (, ) 的字符串。<br>有效的括号对字符串必须满足:<br>1. 每个左括号的右侧都有一个与之匹配的唯一对应的右括号。<br>2. 不同位置的左括号与不同位置的右括号相匹配。<br><br>有效的括号对字符串:<br>``` + '\n' + '('* $n$ + ')'* $n$[6] | Given a string containing only (, ).<br>A valid bracket pair string must satisfy:<br>1. The right side of each left bracket has a unique corresponding right bracket that matches it.<br>2. Left brackets in different positions match right brackets in different positions.<br><br>A valid bracket pair string:<br>``` + '\n' + '('* $n$ + ')'* $n$ |

Table 2: LLMs completion brackets for string prompts

[6] Concatenate a newline character and the string $l^n r^n$ at the end of the cue word.

### 3.3 Validating the DSL Generation Performance of Language Models

By providing language models with prompt words and $l^n r^n \in L(G)$, allowing them to generate the next token $x$, and calculating the probability of $x = l$ or $x = r$, we can obtain the performance of the language model on the task of generating valid bracket pairs. Figure 6 demonstrates the experiment conducted on four models, showing the change in the probability of generating $r$ and invalid strings as $|s|$ increases.

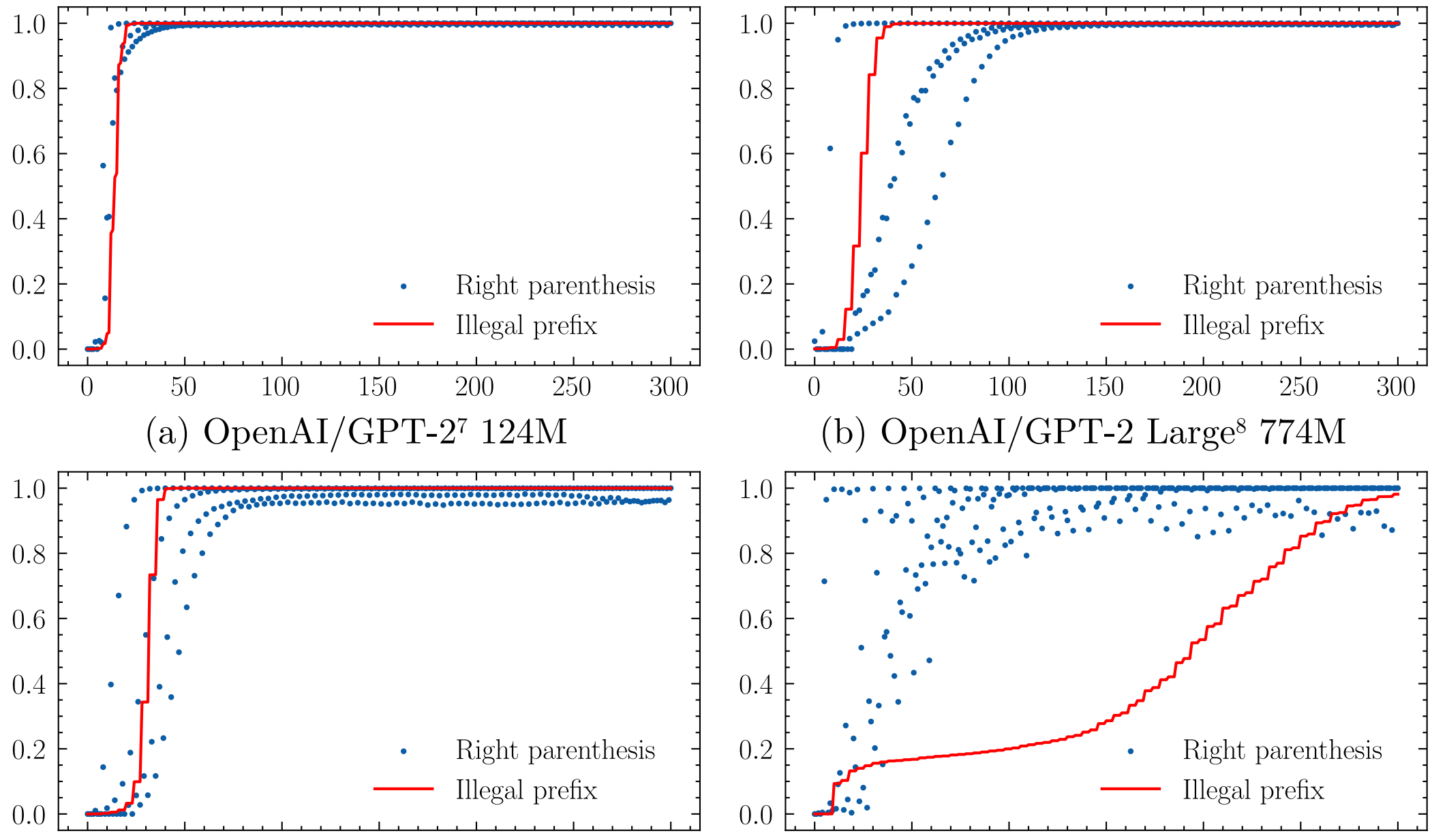

(a) OpenAI/GPT-2[7] 124M (b) OpenAI/GPT-2 Large[8] 774M

(c) OpenAI/GPT-2 XL[9] 1558M (d) Google/Gemma-7B[10] 7751M

Figure 6: Probability of erroneous DSLs increases with model generation length

The data from Figure 6 shows that as the length $|s|$ of the string increases, the error rate $r(e)$ of the four models continuously rises. In other words, under the basic DSL generation task of matching bracket pairs, the DSL generation performance is rapidly deteriorating.

| **Model Names** | **Parameters** | **Date**$^{\text{GMT+8}}$ | $\lvert s\rvert_{r(e)>50\%}$ | $\lvert s\rvert_{r(e)>95\%}$ |
|---|---|---|---|---|
| GPT-2 | 124M | 2019-2-14 | 14 | 20 |
| GPT-2 Large | 774M | 2019-2-14 | 24 | 32 |
| GPT-2 XL | 1558M | 2019-2-14 | 32 | 36 |
| Gemma | 7751M | 2024-2-21 | 194 | 282 |

Table 3: Error rate $r(e)$ versus average string length $|s|$

Table 3 showed that the error rate of GPT-2 with a parameter count of one hundred million rises to 95% when generating strings with a length of 20 characters, and even for Gemma (Team et al. 2024) with a parameter count of nearly 7.8 billion, when the generated string length $|s|$ reaches 282, the error rate exceeds 95%. The performance of these language models in DSL generation is unacceptable.

[7]https://huggingface.co/openai-community/gpt2
[8]https://huggingface.co/openai-community/gpt2-large
[9]https://huggingface.co/openai-community/gpt2-xl
[10]https://huggingface.co/google/gemma-7b

## 3.4 Generating DSLs with Language Models is Challenging

Neural networks (NNs) can work properly within a certain task length, but NN failure with increasing length is a very typical occurrence. While situations where humans can “see at a glance” can be addressed by NNs, tasks that require a more “algorithmic” approach (such as explicitly calculating brackets to check for their matching) are often unreliable due to the “shallow computation” of NNs. Even the current comprehensive ChatGPT with 175 billion parameters finds it difficult to correctly match bracket pairs in long sequences (Wolfram 2023).

LLMs typically use autoregressive models to generate text. In autoregressive models, the probability distribution of the next word depends on the previous words. However, if the model always chooses the word with the highest probability, it will be overly “conservative” and lack creativity and diversity. In industrial practice with LLMs, different samplers can be used to help improve their performance. Common sampling methods include random sampling, greedy search, beam search, and nucleus sampling.

LLMs can also undergo post-processing operations such as “temperature” adjustment before sampling to alter the probability distribution in the logits tensor of the model output layer, thereby affecting the sampling process (Wang, Liu, and Awadallah 2023; Achiam et al. 2023; Touvron et al. 2023). Different sampling parameters are hyperparameters of language models, and these hyperparameters may affect the output and performance of LLMs. The temperature parameter of LLMs is one of the parameters often adjusted in engineering practice. Its value is a floating-point number greater than or equal to 0, typically ranging from $[0, 1]$ (values greater than 1 lead to rapid degradation in model performance) (Renze and Guven 2024), used to adjust the diversity of the model output.

Temperature sampling is related to the Boltzmann distribution, which provides a probability for a system being in a certain state. For instance, in Equation 2, $\rho_i$ describes the probability measure function for the energy and temperature of that state.

$$\rho_i = \frac{1}{Q} e^{-\varepsilon_i/(kT)} = \frac{e^{-\varepsilon_i/(kT)}}{\sum_{j=1}^{M} e^{-\varepsilon_j/(kT)}} \tag{2}$$

$$\operatorname{softmax}\left(-\frac{\boldsymbol{z}}{T}\right)_i = \frac{e^{-z_i/T}}{\sum_{j=1}^{K} e^{-z_j/T}} \tag{3}$$

Equation 2 is analogous to Equation 3′s softmax function, except that the softmax function does not divide by $-kT$ on vector $z_i$ . As depicted in Figure 7, the distribution of $\rho_i$ becomes more uniform as the temperature $T$ increases, while it becomes more concentrated as $T$ decreases. Temperature sampling adjusts the parameter $T$ to alter the probability distribution of tokens, increasing $T$ significantly enhances the diversity of content output by LLMs.

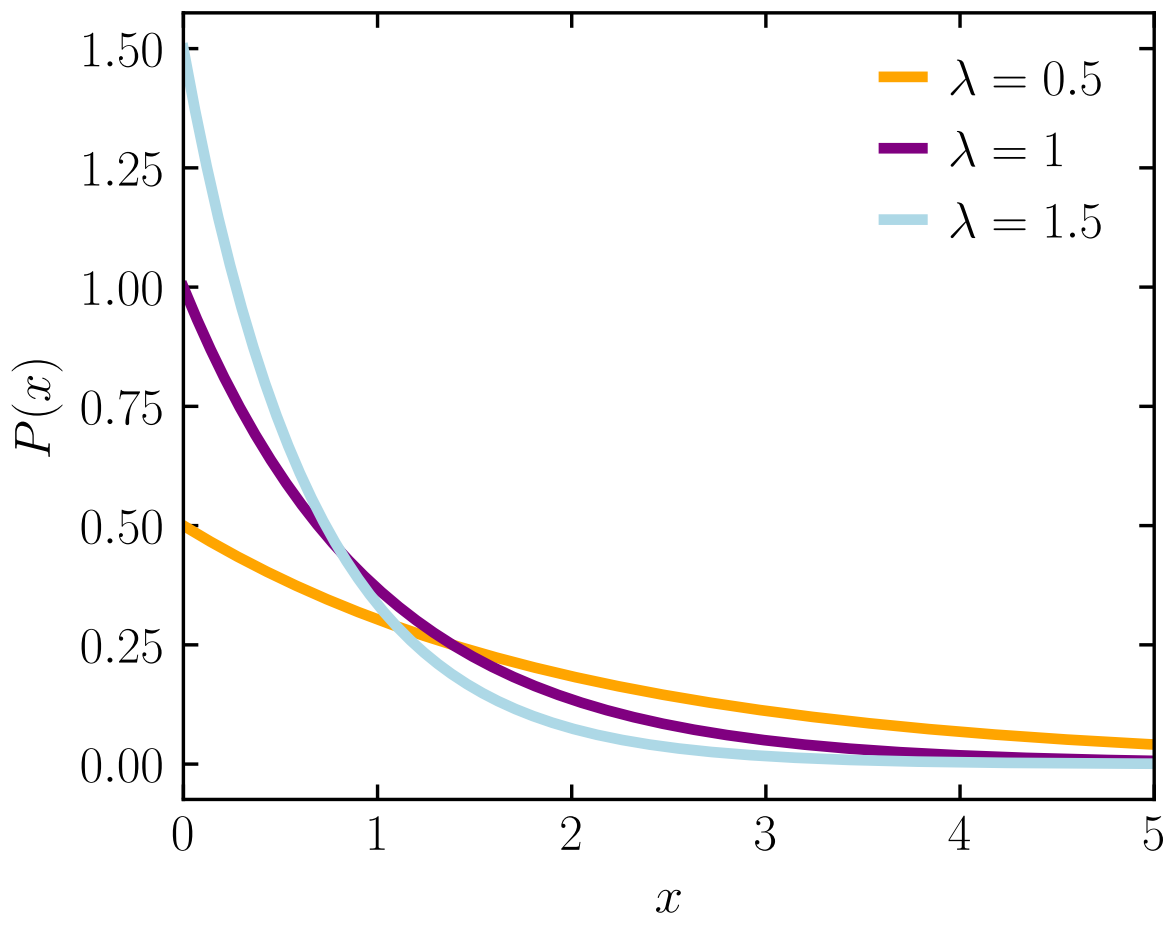

Figure 7: Probability density function plot of exponential distribution (Newystats 2019)

The Boltzmann distribution is an exponential distribution of the form $P(x) \sim \text{Exp}(x)$. As shown in Figure 7, as $\lambda$ decreases from 1.5 to 0.5, the distribution becomes flatter, indicating that increasing temperature $T$ makes other possibilities of words more likely to be chosen. Particularly, at temperature $T = 0$, it is equivalent to selecting the word with the highest probability, i.e., Top-1 token.

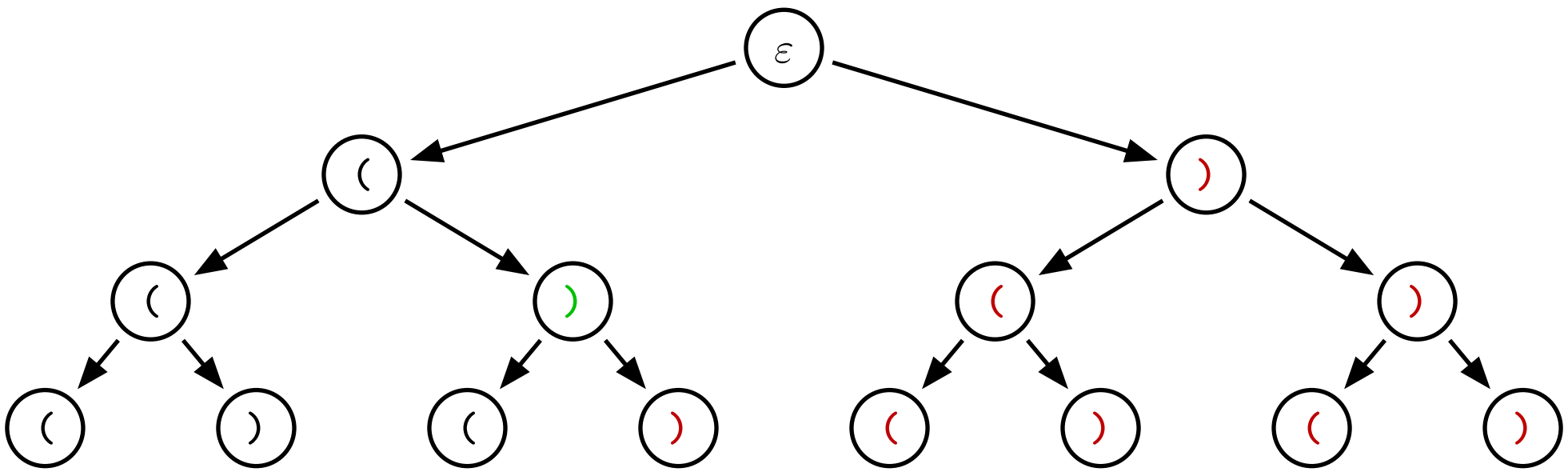

Figure 8: Tree diagram of legal brackets for strings

The autoregressive language model generates a sequence of text, which is akin to sampling from a vast vocabulary. For tasks like matching bracket pairs, the vocabulary can be as minimal as left bracket, right bracket, and the end-of-sequence (EOS) token. Navigating through the tree structure, akin to that illustrated by Figure 8, starting from an empty string ε, and halting at the correct position (generating the EOS token) is challenging. This is due to the extreme scarcity of correct positions, with the number of erroneous paths always surpassing the correct ones. As the sampling in LLMs typically entails stochasticity such as "temperature," a misstep often nullifies previous progress.

## 3.5 Summary of This Chapter

This chapter examined the typical syntactic feature of DSLs, "matching bracket pairs," and provided validation algorithms and patterns for them. Subsequently, the experimental approach to evaluating the performance of language models in generating bracket pairs was introduced. The chapter primarily showcased the performance of four language models in the bracket pair generation task, demonstrating that these language models struggle with DSL tasks. Finally, a theoretical explanation for the suboptimal performance of autoregressive language models in DSL generation tasks was briefly outlined.

# 4 Rules of Generating DSLs, Asynchronous and Coroutine

## 4.1 Levels and Basic Concepts of Formal Languages.

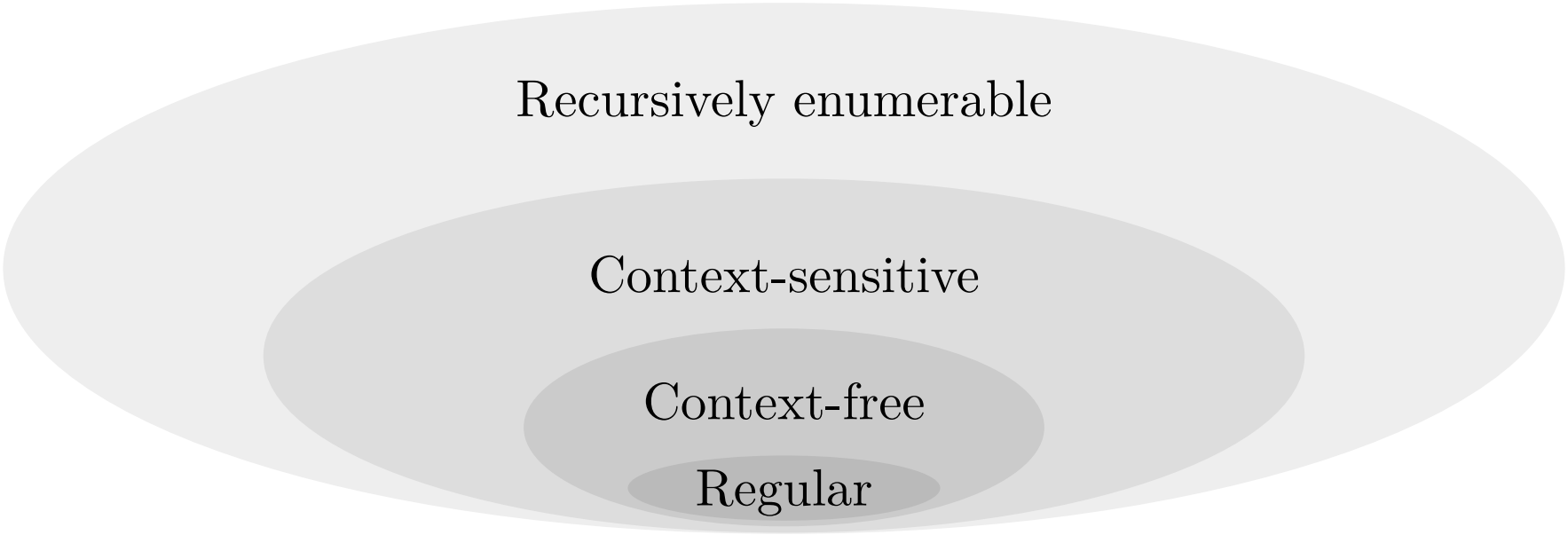

Figure 9: Chomsky genealogy (Chomsky 1956)

Linguist Avram Noam Chomsky proposed the famous Chomsky genealogy in 1956. Chomsky's genealogy is a classification genealogy in computer science that describes the expressive ability of formal grammars. It divides the formal language into four levels (Chomsky 1956), and contains them step by step, as shown by Figure 9 and Table 4.

| Grammar | Languages | Recognizing Automatons | Production Rules |
|---|---|---|---|
| Type-0 | Recursively enumerable | Turing machine | $(\gamma \neq \varepsilon) \to \alpha$ |
| Type-1 | Context-sensitive | Linear-bounded non-deterministic Turing machine | $\alpha A\beta \to \alpha\gamma\beta$ |
| Type-2 | Context-free | Non-deterministic pushdown automaton | $A \to \alpha$ |
| Type-3 | Regular | Finite-state automaton | $A \to aB$ |

Table 4: Four Grammatical Types on Chomsky's Genealogy

Theories that specialize in the study and expression of language grammar are called formal language theories, and languages that can be accurately processed by mathematics or computers are called formal languages. As a formal language, domain-specific language also contains two parts: syntax and semantics. First of all, grammar is the basis of DSL. If there is a problem with the grammar of a string, the string cannot be recognized by the agreed computer program, and it is difficult to simply repair or restore the information.

The formal language is defined on a specific alphabet $\Sigma$. For example, the set {`'('`, `')'`, `&`} is an alphabet (where `&` for stop generating), and the set $\{1, 0\}$ can also be an alphabet. The alphabet of a formal language happens to correspond to the vocabulary of a LLM.

If there is an alphabet $\Sigma$, a string of length $n$ in the alphabet is denoted as $\Sigma^n$, and a string of any length is denoted as $\Sigma^*$. In particular, $\Sigma^0$ is defined to contain only empty strings, the set $\{\varepsilon\}$ . Language $L$ on alphabet $\Sigma$ has syntax $G$ represented as $L(G) \subseteq \Sigma^*$ . If $a, b \in \Sigma^*$, $ab$ means they are connected left and right, $a^k$ means $a$ is repeated $k$ times.

Context-free grammars (CFG) in the Chomskyan lineage of Table 4 are expressive enough to express the vast majority of programming languages and are the preferred form of describing the syntax of programming languages (Mascarenhas, Medeiros, and Ierusalimschy 2014).

## 4.2 The Grammar of a Context-Free Language

The syntax of a formal language is defined by production rules, which consist of a finite set of non-terminal symbols $N$, a finite set of terminal symbols (alphabet) $\Sigma$, a finite set of production rules $P$ in the form $(\Sigma \cup N)^* N (\Sigma \cup N)^* \rightarrow (\Sigma \cup N)^*$, and a start symbol $S \in N$. The quadruple $G = (N, \Sigma, P, S)$ defines a formal grammar.

If $l =$ `'('` and $r =$ `')'`, then a legal sequence of bracket pairs can be represented by Equation 4.

$$P = \{S \rightarrow \varepsilon, S \rightarrow lSrS\} \tag{4}$$

Equation 4 expresses the same meaning as Equation 5, but Equation 5 may be clearer.

$$\begin{aligned} \langle\text{Pairs}\rangle &\rightarrow \langle\text{Pair}\rangle \\ \langle\text{Pairs}\rangle &\rightarrow \langle\text{Pair}\rangle\langle\text{Pairs}\rangle \\ \langle\text{Pair}\rangle &\rightarrow (\ \langle\text{Pairs}\rangle\ ) \\ \langle\text{Pair}\rangle &\rightarrow \varepsilon \end{aligned} \tag{5}$$

Context-free grammars (CFGs) serve as the preferred form for defining programming language syntax and can indeed be employed to define C language. The left side of Figure 10 displays the syntax of a subset of C language defined by CFG production rules. A simple string of C code `if ( x > 9 ) { x = 0; y = y + 1; }` can be decomposed into a tree-like syntactic structure as shown on the right side of Figure 10.

```
    ⟨Stmt⟩ → ⟨Id⟩ = ⟨Expr⟩ ;
    ⟨Stmt⟩ → { ⟨StmtList⟩ }
    ⟨Stmt⟩ → if ( ⟨Expr⟩ ) ⟨Stmt⟩
⟨StmtList⟩ → ⟨Stmt⟩
⟨StmtList⟩ → ⟨StmtList⟩ ⟨Stmt⟩
    ⟨Expr⟩ → ⟨Id⟩
    ⟨Expr⟩ → ⟨Num⟩
    ⟨Expr⟩ → ⟨Expr⟩ ⟨Optr⟩ ⟨Expr⟩
      ⟨Id⟩ → x
      ⟨Id⟩ → y
     ⟨Num⟩ → 0
     ⟨Num⟩ → 1
     ⟨Num⟩ → 9
    ⟨Optr⟩ → >
    ⟨Optr⟩ → +
```

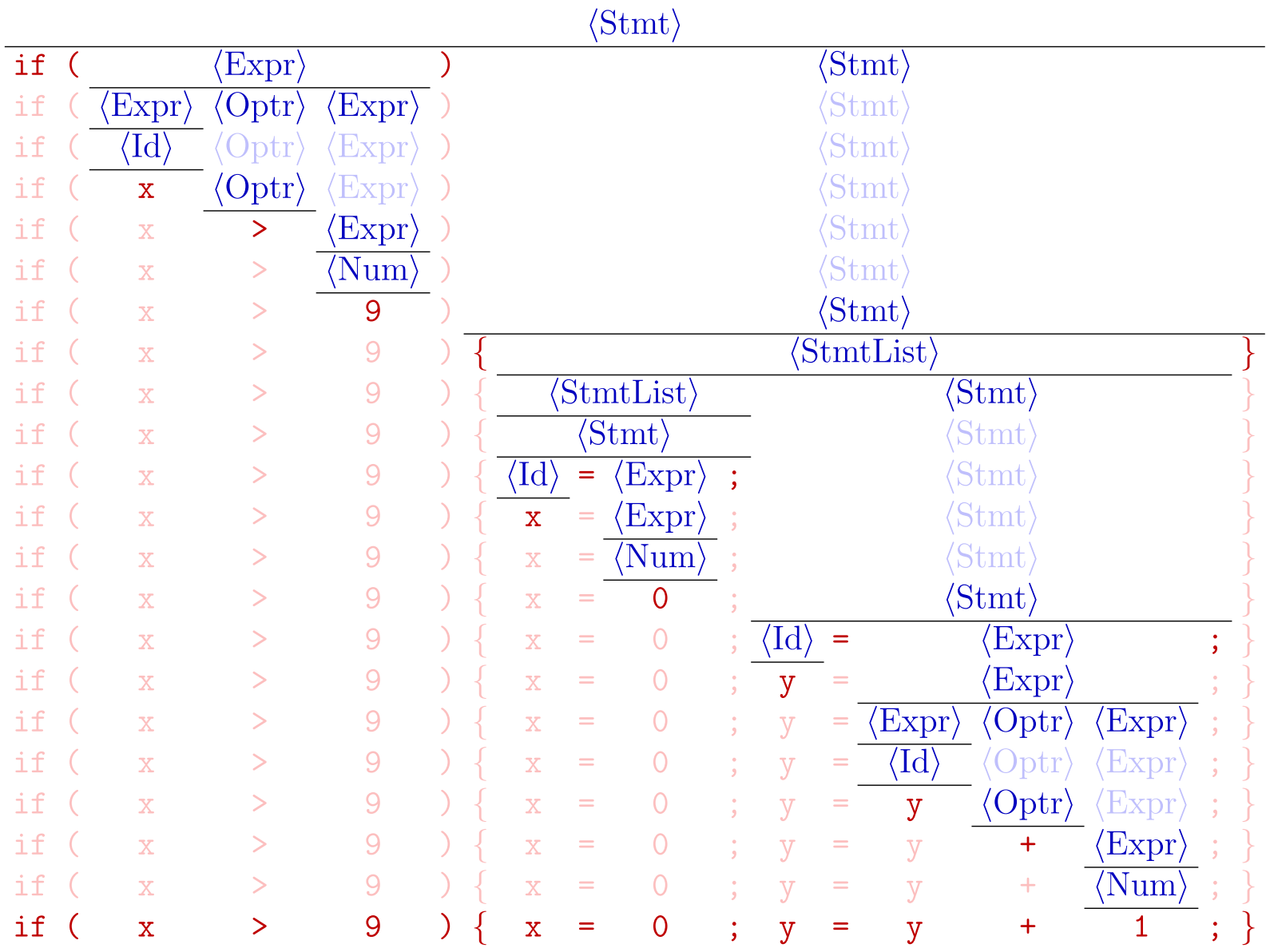

Figure 10: Subset of C language and tree structure (Burghardt 2020)

In Figure 10, ⟨Stmt⟩ denotes a statement, ⟨StmtList⟩ represents multiple statements, ⟨Expr⟩ signifies an expression, ⟨Id⟩ stands for an identifier, ⟨Num⟩ denotes a number, and ⟨Optr⟩ indicates an operator. These non-terminals, expanded according to rules, can constitute a valid C language (Kernighan and Ritchie 1988) code string.

## 4.3 Common Domain-Specific Language Rules

A series of mathematical symbols and extensions used to express formal languages can be referred to as metalanguages, which are the languages used when discussing or studying languages themselves. In addition to the aforementioned mathematical symbols used to express DSLs, Backus-Naur Form (BNF) or its extension EBNF can also be used to express DSLs, and essentially, they are equivalent to the formal language grammar $G = (N, \Sigma, P, S)$ mentioned earlier. BNF is widely used in fields such as programming languages, instruction sets, and communication protocols, including various DSLs described in this paper.

The need to define a formal syntax rule for various internet technology specifications has led to the emergence of another enhanced form of BNF called Augmented BNF (ABNF), which has been widely embraced (Crocker and Overell 2008). ABNF is an extension of BNF that balances compactness, simplicity, and reasonability, making it suitable for describing the syntax of DSLs, and it has become an internet standard (STD 68, RFC5234).

```json
{
  "authors": "Jiaye Wang",
  "keywords": [
    "Large language models",
    "Structured content generation",
    "Computer-aided programming",
    "Constrained decoding",
    "Coroutine",
    "Metalanguage"
  ],
  "date": 2024
}
```

Figure 11: A JSON string example

A typical DSL is the JSON language, exemplified by Figure 11, which is not a regular language due to the presence of bracket pairs that can "recursively" nest within each other, making it a typical context-free language. Some syntax rules of JSON language can be described using a metalanguage as shown in Figure 12.

```
JSON   := [ ws ] value [ ws ]
value  := false | null | true | object | array | number | string
object := '{' [ member *( ',' member ) ] '}'
array  := '[' [ JSON   *( ','   JSON ) ] ']'
number := [ '-' ] int [ frac ] [ exp ]
string := '"' *char '"'
member := [ ws ] string [ws] ':' JSON
exp    := ('e' | 'E') [ '-' | '+' ] 1*digit
int    := '0' | ( onenine *digit )
frac   := '.' 1*digit
```

Figure 12: A brief syntax representation of JSON

In Figure 12, `ws` denotes any whitespace character (e.g., `U+0020`, `U+000A`, `U+000D`, `U+0009`, etc.), `char` denotes any character (escaping if ambiguous), `digit` denotes a digit (0 to 9), and `onenine` denotes a non-zero (1 to 9) digit character. Additionally, a string in the form `[ A ]` indicates that `A` is optional (i.e., `A` or empty string $\varepsilon$); `*A` denotes that `A` can occur zero or more times; `1*A` denotes that `A` must occur at least once; `A | B` denotes either `A` or `B`; strings enclosed in single quotes `'A'` represent specific terminal symbols belonging to $\Sigma$.

```
request-header := Accept
                | Accept-Charset
                | Accept-Encoding
                | Accept-Language
                | Authorization
                | Expect
                | From
                | Host
                | If-Match
                | If-Modified-Since
                | If-None-Match
                | If-Range
                | If-Unmodified-Since
                | Max-Forwards
                | Proxy-Authorization
                | Range
                | Referer
                | TE
                | User-Agent
```

Figure 13: Syntax diagram of HTTP/1.1 request header fields

Besides JSON, the Hypertext Transfer Protocol (HTTP) can also be regarded as a DSL since its grammar is defined by ABNF in RFCs such as RFC 2616 (Nielsen et al. 1999). HTTP is an application-level protocol designed for distributed, collaborative, and hypermedia information systems, widely utilized across the global internet.

Elements like `Accept`, `Accept-Charset`, and `Accept-Encoding` in HTTP/1.1 request header fields are part of HTTP, with some of its syntax rules illustrated in RFCs like Figure 13, serving as simple examples.

For document formatting, Typst is a programmable markup language and software designed for typesetting (Mädje 2022). Figure 14 demonstrated how Typst renders its DSL into rich text for user presentation. Similarly, the DSL of Typst can also be expressed by BNF, with partial syntax rules of Figure 15.

```
#set par(leading: 0.5em)
- *Sine Function:*
  - $f(x)=sin(x)$
- *Cosine Function:*
  - $f(x)=cos(x)$
```

a. Typst DSL sample

- **Sine Function:**
  - $f(x) = \sin(x)$
- **Cosine Function:**
  - $f(x) = \cos(x)$

b. Processed rich text

Figure 14: Typst renders DSL to rich text

```
linebreak   := '\' '+'?
text        := (!special)+
escape      := '\' special
quote       := "'" | '"'
strong      := '*' markup '*'
emph        := '_' markup '_'
raw         := '`' (raw | .*) '`'
link        := 'http' 's'? '://' (!space)*
math        := ('$' .* '$') | ('$[' .* ']$')
heading     := '='+ space markup
list        := '-' space markup
enum        := digit* '.' space markup
desc        := '/' space markup ':' space markup
label       := '<' ident '>'
ref         := '@' ident
markup-expr := block | ('#' hash-expr)
```

Figure 15: Basic syntax of Typst (Mädje 2022)

## 4.4 Recursive Descent Parser and Generator

To enable computer programs to recognize DSLs, it is necessary to analyze each substring $s$ in $L(G)$ as symbols first, and then generate a syntax tree according to the production rules $P$. Recursive descent is a common parsing method, which is a top-down parsing method that recursively parses from the root node down to the leaf nodes. Each non-terminal $n \in N$ of a recursive descent parser corresponds to a recursively callable function.

Formal grammars containing left recursion cannot be parsed by simple recursive descent parsers unless they are transformed into weakly equivalent right-recursive form. Some studies suggest that by using "reduction" left-recursive grammars (along with all other forms of general CFGs) can be accommodated in more complex top-down parsers (Frost and Hafiz 2006).

```
value = async function(set: Setter<JSONValue>) {
    const token = await this.token.read();
    switch (token.type) {
        case TokenType.LeftBrace:
            await this.token.unread(token);
            await this.object(set);
            break;
        case TokenType.LeftBracket:
            await this.token.unread(token);
            await this.array(set);
            break;
        case TokenType.Digits:
        case TokenType.Negtive:
        case TokenType.Positive:
            await this.token.unread(token);
            await this.number(set);
            break;
        case TokenType.Bool:
        case TokenType.Null:
        case TokenType.String:
            await set(token.value);
            break;
        default:
            console.error(`unexpected token: ${token}`);
    }
}
```

Figure 16: Parsing JSON values using asynchronous recursive descent

As shown in Figure 12, the syntax of JSON corresponds to a context-free language and can naturally be parsed into data structures within applications using recursive descent parsing techniques. Figure 16 presents an example of an asynchronous recursive descent parser (based on the TypeScript language, but to be considered as pseudocode).

```
number = async function(set: Setter<number>) {
    // sign part (optional)
    const sign = await this.sign();
    // integer part
    const intToken = await this.matchToken(TokenType.Digits);
    let num = parseInt(intToken.value as string);
    // fraction part (optional)
    const dotToken = await this.matchOptionalToken(TokenType.Dot);
    if (dotToken !== undefined) {
        const intToken = await this.matchToken(TokenType.Digits);
        num += parseFloat(`0.${intToken.value}`);
    }
    // exponent part (optional)
    const eToken = await this.matchOptionalToken(TokenType.E);
    if (eToken !== undefined) {
        // sign part (optional)
        const sign = await this.sign();
        // exponent part
        const intToken = await this.matchToken(TokenType.Digits);
        const int = parseInt(intToken.value as string);
        num *= Math.pow(10, sign * int);
    }
    await set(sign * num);
}
```

Figure 17: Parsing JSON numbers using asynchronous recursive descent

Using the recursive descent parsing approach typically requires some utility functions, such as the token reading function shown in Figure 16 `token.read()` and the function for token backtracking `token.unread()`. This is akin to having a pointer scanning through the string to be parsed from start to finish, with the ability to backtrack when necessary.

By reversing the top-down parser, starting from the start symbol $S$, and recursively generating terminal symbols $x$ from the production rules $P$ of the grammar $G$, a recursive descent generator can be obtained. Choosing different recursive function call paths in each generation function can be seen as sampling in the language $L(G)$ until reaching a leaf node (symbols like EOF are also considered leaf nodes), thus completing language generation.

## 4.5 Asynchronous, Coroutine, and Generator Functions

Top-down parsers have a typical issue: usually, before parsing a string $s$ in $\Sigma^*$, the string $s$ should already be fully loaded into memory. The problem with this approach arises if $s \notin L(G)$ initially or if the data gets corrupted during transmission, the parser still needs to wait for the entire transmission to complete before parsing, inevitably resulting in parsing failure. For instance, if a 1024GiB super-large JSON data file is transmitted and parsing fails, significant computational costs before parsing are clearly wasted.

The same principle applies to LLMs. If a language model outputs a sequence of 3000 tokens, and then it is discovered by the subsequent program that the DSL is incorrect, the information cannot be recognized, thus rendering a large amount of neural network computation "wasted". One solution to this problem is to adopt asynchronous and coroutine techniques.

Asynchronous refers to multiple operations that can occur simultaneously without waiting for each other to complete (blocking) (Davies 2012). This can be achieved in various ways, such as multithreading, multiprocessing, or event-driven programming. Coroutines are a type of asynchronous computer program component that is a function capable of pausing and resuming execution. Coroutines are commonly used to implement asynchronous programming because they allow multiple operations to be executed in one thread without blocking other operations.

Generators (also referred to as "semi-coroutines") (Dahl 1972) are a subset of coroutines. Although both can pause execution through multiple yield expressions and allow re-entry at multiple entry points, their main difference lies in the ability of coroutines to control the location where the program resumes execution after yielding, while generators cannot. Generators can only transfer control back to their caller (Foundation 2019). In other words, since generator functions are primarily used to simplify the writing of iterators, their yield statements are not intended for jumping to another coroutine but rather for passing a value back to the calling parent routine.

Generator functions supported by various high-level languages are one way to implement coroutines, as they can generate a sequence of values. Generator functions use the yield keyword to pause execution (while preserving the current execution context) and produce a value. Taking TypeScript as an example, a generator `g` can resume execution using the `g.next()` function, end execution using the `g.return()` function, and throw exceptions using the `g.throw()` function.

Similar to the `async` and `await` keywords shown by Figure 17, `async` represents a function is asynchronous (thus its call returns a generator object), and `await` represents waiting for an asynchronous operation to complete. Such asynchronous generator functions can be used for parsing and generating DSLs since they can pause and resume execution during parsing or generation without blocking other operations.

For example, when we want to parse a number like `-123.456e-7`, we can implement it using an asynchronous generator function, as shown by Figure 17. First, parse the sign part `'-'`, then parse the integer part `'123'`, next encounter the decimal point `'.'` to parse the decimal part `'456'`, and finally encounter `'e'` to parse the sign part of the exponent `'-'` and the integer part of the exponent `'7'`. After each parsing step is completed, we can pause the parser, wait for other operations to finish, and then resume the parser to continue parsing.

By employing an asynchronous recursive descent parser alongside an asynchronous string reader (which constructs a buffer for loading strings), these two modules can collaborate to form a coroutine-based apparatus. Such a parser can load strings while parsing DSLs into specified data structures and can immediately stop loading and parsing when encountering an error, thus saving computational resources.

## 4.6 Generating Domain-Specific Languages with Coroutines

The implementation of coroutines does not depend on any specific programming language features. Although some programming languages do not support the `yield` keyword, any Turing machine can equivalently implement yield because yield can be converted to "switch-case" and the storage of context states.

In fact, any `async` and `await` can be equivalently transformed into generator functions and `yield`. This means that asynchronous operations can be based on coroutines. Therefore, asynchronous recursive descent parsers can also be equivalently transformed into recursive descent parsers based on generator functions.

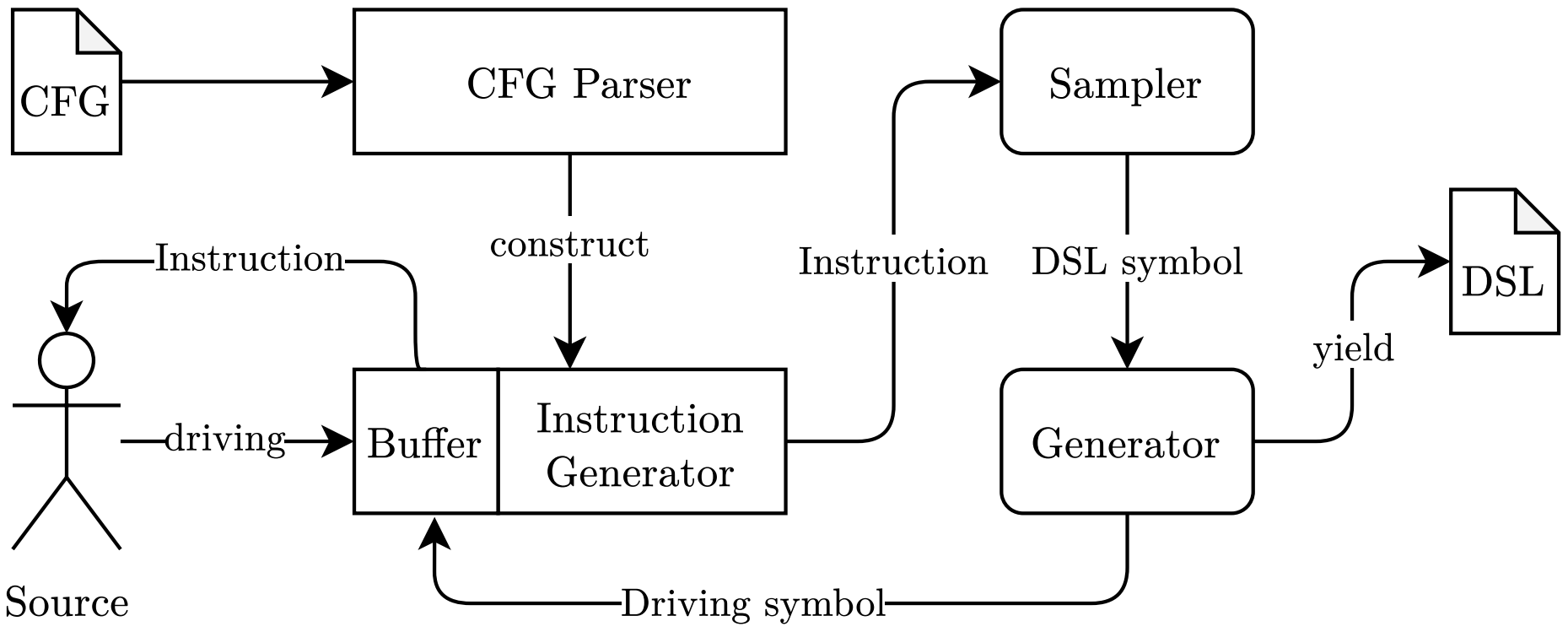

Figure 18: A module for generating DSL based on asynchronous

The coroutine module designed in this paper, as shown in Figure 18, accepts input streams from users or agents and is driven to maintain the parsing and generation states of the DSL internally through asynchronous and coroutine mechanisms. Each drive to the module is called a query (which can carry information or be an empty query), and each query to the module returns a guidance instruction. Guidance instructions can carry sets of symbols describing the expected generation, precise error messages and recovery information, read/write operation information, and information about the DSL prefix. Users and/or agents accept guidance instructions to continue driving the module until EOF is input. Throughout this process, the module asynchronously parses the input stream into the agreed data structure to ensure correct handling by the application.

This module accepts a pre-defined CFG from developers and constructs an asynchronous guidance instruction generator through the CFG parser of this module. This generator, driven by input sources through a buffer, produces guidance instructions (Instruction DSL) feedback to the input source. The input source can also be a sampler and generator. The sampler can convert guidance instructions into a DSL symbol through random, custom logic, or neural networks, and hand it over to the generator to generate DSL while driving the asynchronous instruction generator to continue generating guidance instructions until receiving the EOF instruction, completing the task of generating domain-specific languages.

The key points of this module lie in the adoption of "asynchronous mechanisms" and "instruction generation." Specifically, the first technical key point of this module is that the asynchronous language guidance and generation mechanism adopts `yield*` (yield from). The second technical key point of this module is to use an asynchronous generator module to generate instructions for generating DSL.

```
let input = yield new Instruction(...args);
```

Figure 19: Pseudocode that produce instructions for generating DSL

Figure 19 described herein generates Instruction DSL through yield to guide DSL verification, parsing, and generation. Here, args represent the information carried by the instruction, and input denotes the information passed from program B (DSL verifier, parser, and generator) to program A upon receiving the instruction. Yield allows program A (DSL parser) to immediately return an instruction generator to program B.

```
let instruction = generator.next(...inputs);
```

Figure 20: Pseudocode that receives instructions for generating DSL

In the Figure 20, Program A pauses midway through execution. It then awaits Program B to trigger the instruction generator's next function, resuming Program A's execution. Simultaneously, Program A obtains the query, executes to the yield point, and generates the guiding instruction. Program B obtains the guiding instruction from the return value of the next function. This process repeats until Program B continuously triggers Program A to complete DSL analysis-related algorithms.

The "yield*" delegates the generation of DSL guiding instructions to another program. "yield*" is also a high-level feature of some programming languages. Through "yield*", the form of CFG for DSL can be modularly expressed, and recursive algorithms can be used to process or generate DSL. The top function that starts this module is referred to as the top level. If this module accepts the CFG of JSON, then `let top = yield* json` indicates that the top-level module generates JSON.

```
let json = yield* select(
    number,
    string,
    object,
    array,
    false,
    true,
    null
)
```

Figure 21: Generate CFG statements for generating JSON

Figure 21 indicates that JSON consists of one of the following: a number, a string, a array, an object, true, false, or null. Similarly, the specific grammar for number, string, array, etc., can be determined based on the CFG. The figure above illustrates that false, true, and null are terminal symbols because they are composed of fixed strings `'false'`, `'true'` and `'null'` respectively. The generation process for objects and arrays can produce numbers, strings, false, true, null, and even themselves recursively to form a recursive definition. Since the yield* mechanism can delegate asynchronous tasks, recursive definitions can be implemented. This module can generate equivalent asynchronous programs based on this or any CFG.

The other key point of this framework lies in the asynchronous generation of Instruction DSLs, which serve as the guidance for validating, parsing, and generating DSLs. They can also provide precise error information and error recovery information based on asynchronous inputs. The minimal unit of information on specific tasks is referred to as a token, which can be, for example, a bit, byte, character, number, especially characters or words. Specifically, the Instruction DSL, as demonstrated by Table 5, consists of control instructions, constraint instructions, suggestion instructions, and symbol instructions.

| Instruction Types | Instruction Examples |
|---|---|
| Control instructions | Instructions to view the next N tokens, instructions to read the next N tokens, instructions to cancel the first N reads, instructions to generate certain tokens and EOF instructions, etc. |
| Restriction instructions | Instructions that limit the next reading of a specific token, limit the next reading of tokens to only certain specific optional instructions, and limit the next reading of tokens to a certain range of instructions and restrictions. The token read at one time cannot be some specific instructions, the instruction that restricts the token read next time should not be in a certain range, etc. |
| Suggested instructions | Instructions that read the penultimate N token that do not conform to the syntax, instructions that suggest the specific tokens to be read next, instructions that suggest the range of the tokens that should be read next, etc. |
| Symbol instructions | Instructions to enter a certain non-terminal symbol $m \in N$, instructions to leave a certain non-terminal symbol, instructions to complete the generation of a certain terminal symbol $x \in \Sigma$, etc. |

Table 5: Examples of instructions for generating DSL

## 4.7 Demo of the Coroutine Module that Generates DSLs

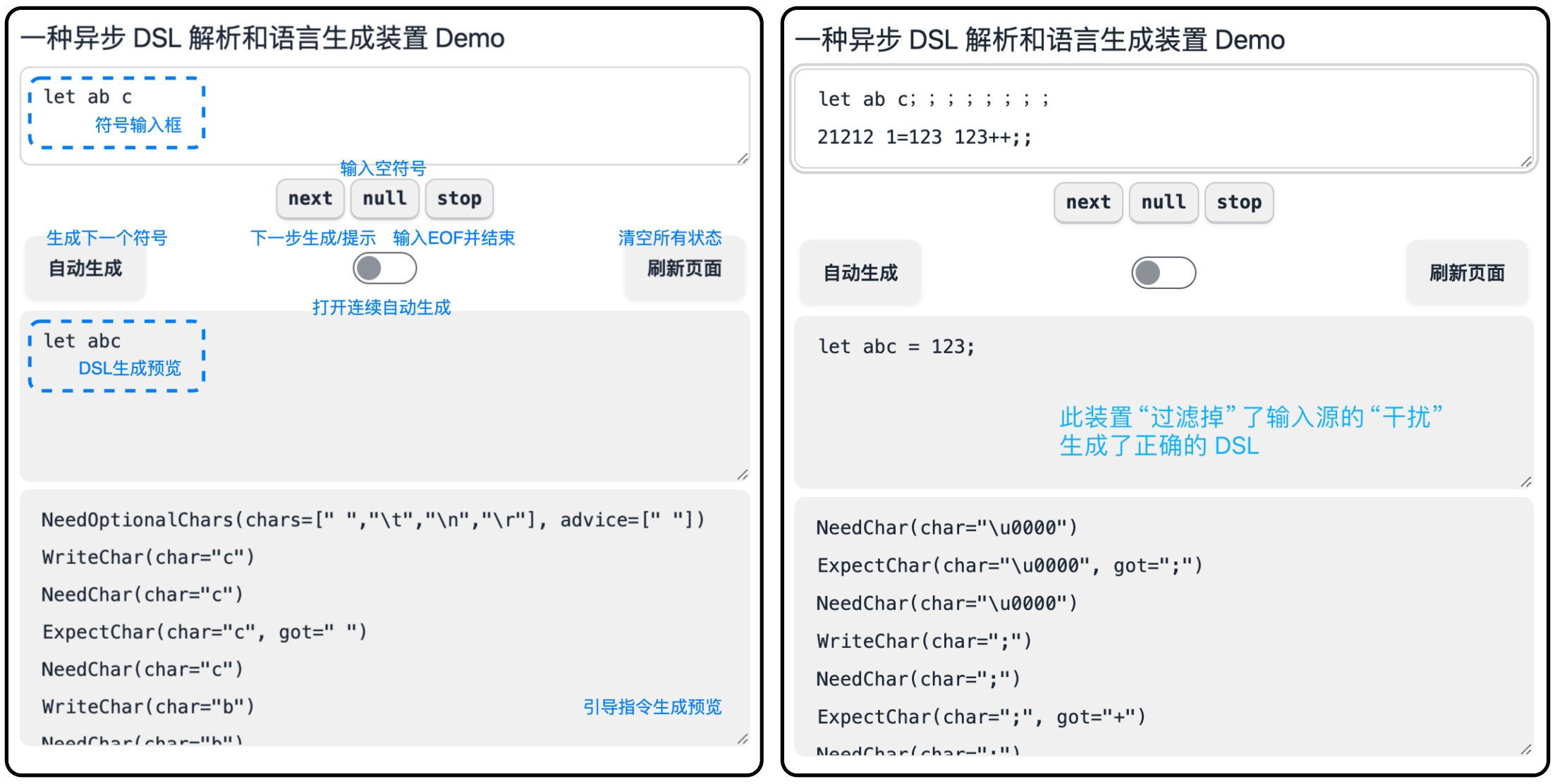

(a) Demo of DSL generation instructions (b) Demonstrated on a simple DSL

Figure 22: An interactive demo using the module in Section 4.6

The Figure 22 is an interactive demo preview interface deployed on the Section 4.6 module. This demo provides a human-computer interaction interface for manually operating this module. For ease of understanding, this demo is configured with a simple grammar CFG, which is a simple grammar like `let abc = 123;` . The interaction mainly consists of three parts: symbol input box, control buttons, DSL generation preview, and guided instruction generation preview. The symbol input box is an input source that can collect symbols entered via the keyboard and then input them into the buffer of this module. Control buttons can input control symbols that cannot be entered by the symbol input box, such as next, null, and stop. Clicking the auto-generate button, this module can generate the next symbol once. If the auto-generate switch is turned on, it will continue generating until the DSL end symbol without any other input. The refresh page button can reset this module and this interaction interface. The DSL generation preview area displays the legal DSL prefixes collected by this module from the input source. The guided instruction generation preview area displays the guided DSL generation instructions generated by this module based on the input source (input box or sampler), which can further guide the generation module to generate new legal DSL prefixes in the DSL preview area.

An asynchronous, coroutine-based, robust DSL generation and parsing module can be used in information transmission, integrated development, educational software, data mining, and other fields. The Table 6 introduces the application of the Section 4.6 module.

| Fields | Introductions to Possible Uses |
|---|---|
| Information Transmission | This module can asynchronously parse DSL into data structures and drive other programs to run with higher robustness. There is no need to wait for the entire transmission to complete before parsing, nor will it cause errors and information loss due to errors in part of the DSL structure of the incoming message. |
| Integrated Development | This module will be able to respond to user input with higher efficiency and provide real-time feedback of boot information or error information generated by the DSL to help developers input. |
| Educational Software | This module provides precise spelling suggestions, error feedback, and fix suggestions. Therefore, it is suitable as a DSL learning guide to help students quickly master the grammar of DSL. |
| Data Mining | This module is able to scan characters one by one from a large amount of text and extract CFG-compliant strings to find the target DSL from a large amount of data. |

Table 6: Possible uses of coroutine-based DSL generation module

## 4.8 Summary of This Chapter

First, this chapter introduces the hierarchy and basic concepts of formal languages, presents the relationship between context-free languages and productions, and then describes the use of metalanguages such as BNF to represent syntax, listing common examples of domain-specific languages and their brief syntactic rules. Subsequently, this chapter introduces recursive descent parsers and generators, highlighting the potential computational resource wastage problem due to the lack of asynchronous support in common parsers. It also explores the relationship between asynchronous operations, coroutines, and generator functions (denoted as “asynchronous” $\subset$ “generator functions” $\subset$ “coroutines”), and proposes a DSL generation module based on “yield*” to address the aforementioned issue. Finally, this chapter demonstrates the interaction of a coroutine-based DSL generation module and its potential applications.

# 5 Generating DSLs via Language Model Sampling

After obtaining a coroutine-based language generation or parsing module, an input source is still required. The input source can be humans or specific computer programs. This chapter will focus on using autoregressive language models as input sources and employing constraint modules to restrict the sampling of language models to generate DSLs expected by developers.

### 5.1.1 Autoregressive Language Models

In fields such as statistics, econometrics, and signal processing, the autoregressive model is a type of model used to describe stochastic processes. It can be employed to describe certain natural processes that change over time. The autoregressive model $AR(p)$ can be described by Equation Equation 6, where $X_t$ is the random variable at time $t$, $p$ is the order of the model, $\varphi_i$ are the coefficients of the model, and $\varepsilon_t$ is the random error at time $t$.

$$X_t = \sum_{i=1}^{p} \varphi_i X_{t-i} + \varepsilon_t \tag{6}$$

Since the advent of the renowned Transformer model in 2017, it has provided inspiration and insights across various domains. Researchers in different fields have derived a series of models based on Transformer, such as those for predicting time series, click-through rates in search recommendations, and user retention prediction. The well-known Transformer is a self-attention-based model (Vaswani et al. 2017), and OpenAI's globally popular ChatGPT, released at the end of 2022, is precisely such a representative model, employing an autoregressive architecture, learning from left to right.

Autoregressive models utilize the preceding context to estimate the probability distribution of terminators in the alphabet $\Sigma$, predicting the next terminator. Given an alphabet $\Sigma$ and an input string $x = (x_1 \cdot x_2 \cdot \ldots \cdot x_n) \in \Sigma^*$, an autoregressive model can compute the probability distribution $p(x_{n+1})$ of the next terminator $x_{n+1}$ as shown in Equation 7. Thus, such autoregressive models are recurrent language models, applicable to generating natural language or various types of tasks such as DSLs.

$$\begin{aligned} p(x) &= \prod_{i=1}^{n} p(x_i \mid x_1 \cdot x_2 \cdot \ldots \cdot x_{i-1}) \\ p(x_{n+1}) &= p(x_{n+1} \mid x_1 \cdot x_2 \cdot \ldots \cdot x_n) \end{aligned} \tag{7}$$

## 5.2 Basic Idea of Language Model Generating DSLs

Next, taking JSON grammar as an example, we illustrate the basic logic of language generation and repair. If the input source has already entered the fragment `{ "key"`, then the coroutine module as mentioned by Section 4.6 can determine that the next character according to the CFG of JSON must be a colon. Why a colon? This requires referencing the JSON syntax defined by the official specification[11] according to the McKeeman Form convention. As shown in Figure 12, the definitions of object, members, and member are presented.

[11] https://www.json.org/json-en.html

```
object  := '{' ws '}'
         | '{' members '}'
members := member
         | member ',' members
member  := ws string ws ':' element
string  := '"' characters '"'
```

Figure 23: Definition of object, members and member in JSON grammar

We always ignore whitespace characters (ws). In the preceding context `{ "key"`, the generation path is object → members → member, and the member has already reached the end of the string. To generate a member, the next character can only be a colon. If the next character in the input source is not a colon, it definitely does not comply with the JSON syntax.

Similarly, in the preceding context `{ "key": 1,`, the next character can only be a double quotation mark, because according to the CFG, the comma at the end implies that the members of this object have not ended yet, and after the comma must be new members, which start with a string, and the beginning of a string must be a double quotation mark.

Taking JSON as an example, in Figure 24, the LLM indicates that the first character of JSON can only be one of the characters in the set {`'{'`, `'['`, `'"'`, `'0'`, `/1-9/`, `'-'`, `'t'`, `'f'`, `'n'`}. To generate JSON, the generator can only select one character from this set as the beginning of the JSON string.

JSON prefix empty string $s$, what is $M_P(s)$? (A)

(B) In the context of JSON (JavaScript Object Notation), we typically consider the syntax rules of JSON to determine which strings or characters can appear in specific contexts. JSON is a lightweight data interchange format based on JavaScript object literal syntax but independent of any language, supported by many programming languages. If we consider an empty string $s$ as a prefix, then we want to identify all possible characters $m$ such that $sm$ is a valid prefix of a JSON structure. In JSON, a structure can be:

- An object, starting with `'{'`
- An array, starting with `'['`
- A string, starting with `'"'`
- A number, starting with any digit or `'-'`
- A boolean value, starting with `'t'` (true) or `'f'` (false)
- A null value, starting with `'n'`

Therefore, for an empty string $s$ , the set $M_P(s)$ will contain all characters that can begin a JSON structure: $M_P(s) =$ {`'{'`, `'['`, `'"'`, `'0'`, `/1-9/`, `'-'`, `'t'`, `'f'`, `'n'`}

This includes all digit characters since a number can begin with any digit, and the symbol `'-'`, as a negative number is also a valid JSON number. Note that this set does not include spaces, tabs, or other whitespace characters, as while they can appear in JSON text, they do not constitute independent JSON structures.

Figure 24: Ask LLMs about possible starting character sets for JSON

When the correct character is entered, for example, if the character '-' is entered, the next character can only be '0' or /1-9/ (this represents a JavaScript regular expression that can only match one of the characters $1, 2, 3, ..., 9$), because the preceding '-' requires the continuation to form a JSON number and so on.

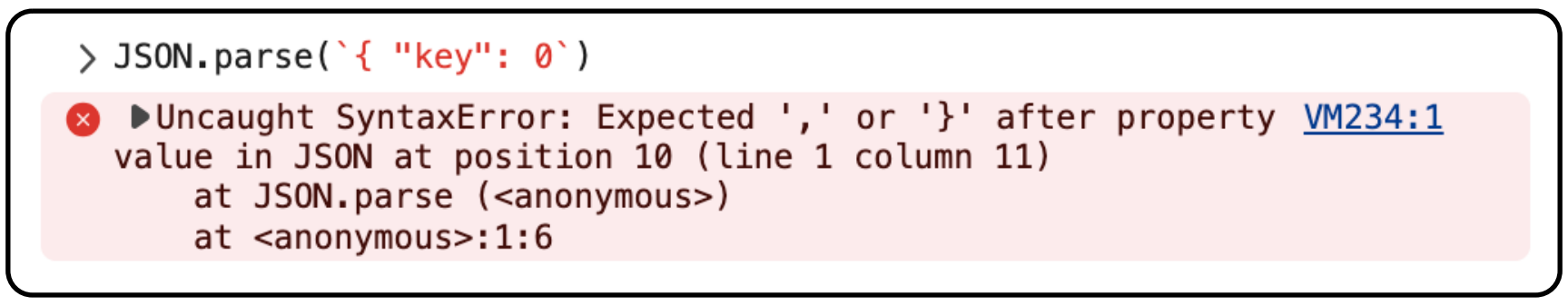

Figure 25: Microsoft Edge browser cannot provide sufficiently precise hints

Taking the V8 JavaScript runtime in the Microsoft Edge browser as an example, as shown in Figure 25, Edge displays the expected characters as only ',' and '}', lacking ['.', 'E', 'e']. Therefore, it cannot provide complete and accurate error and recovery information. Other Parser Generators (such as Antlr, Lark, etc.) also face similar situations, unable to provide error ranges and recovery suggestions precise enough to generate DSLs.

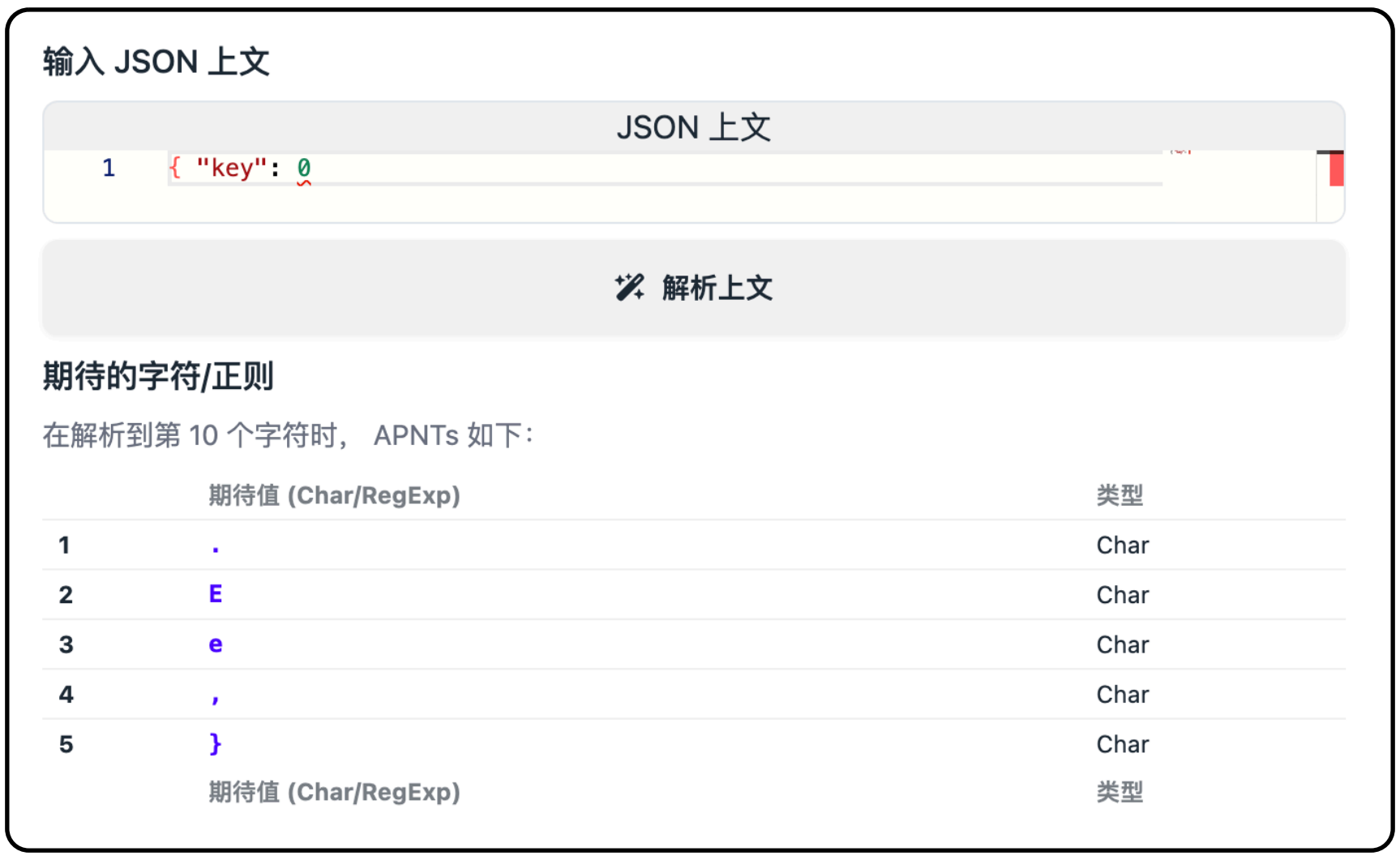

Figure 26: Demonstration of computing all possible next character sets

Figure 26 demonstrates an apparatus in this paper with JSON as a CFG example. When parsing `{ "key": 0` , this apparatus provides more complete and precise information for DSL generation.

Figure 27 provides a schematic diagram that partially expresses the production relationships of symbols in the JSON CFG. It can be observed that the root node of this graph is the `json` node, and there are clearly many cycles in the graph. If we let the language model generate JSON, it is like navigating, selecting routes, and identifying endpoints on such a complex, maze-like graph. This can be cited as a reason why language models struggle to generate longer, legal JSON strings without relying on better decoding apparatus.

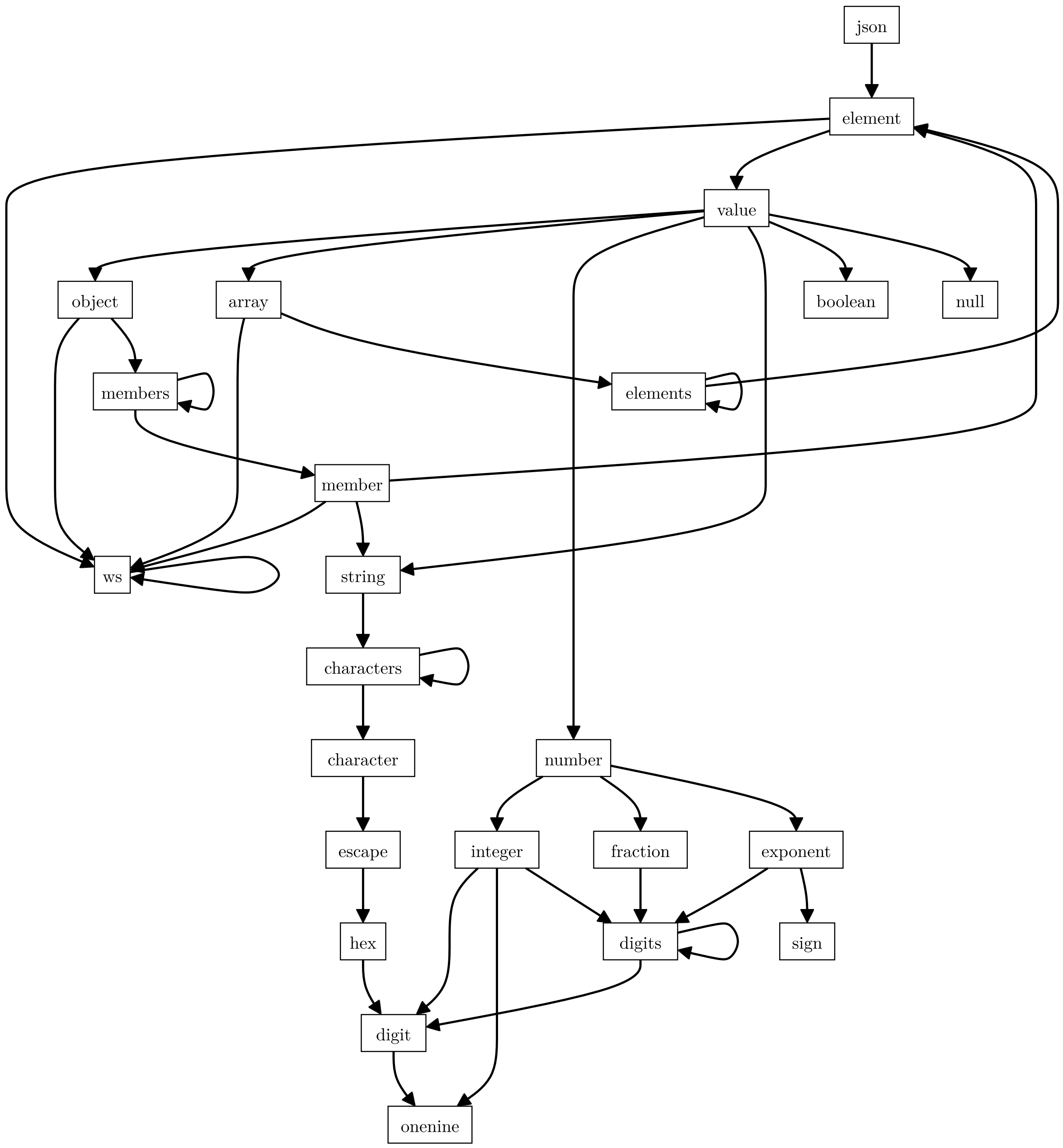

Figure 27: A cyclic diagram of JSON CFG

If we define an alphabet $\Sigma$ , a context-free grammar $G$ and a language $L(G)$ , call $P(G)$ in Equation 8 a prefix language on language $L(G)$ .

$$P(G) = \{x \in \Sigma^* \mid \exists y \in \Sigma^* : xy \in L(G)\} \tag{8}$$

$$M_P(s) = \{m \in \Sigma \mid sm \in P(G)\} \tag{9}$$

If there were a mechanism capable of providing the next valid set of tokens $M_P(s)$ based on a string $s$ in $P(G)$, then the language model could generate a new $s'$ in $P(G)$ by sampling according to the probability distribution $p(x_{n+1})$ over $M_P(s)$ as specified in $M_P(s)$, until the mechanism yields an end symbol from grammar $G$, at which point the model terminates its generation having produced $s'$ in $L(G)$. This constitutes the method for generating DSLs as expected by developers from a language model.

## 5.3 Constrained Decoders for Autoregressive Language Models

Since Torsten Scholak et al. from ElementAI proposed the Parsing-Informed Constraints for Autoregressive Decoding (PICARD) in 2021 (Scholak, Schucher, and Bahdanau 2021), researchers at MIT have adopted an Earley parsing-based algorithm to constrain large models' generation (Wang, Wang, Wang, Cao, A Saurous, and Kim 2024). The basic idea is to devise a mechanism that, given a string $\hat{y}$ generated by a language model, computes a prefix $y_{\text{prefix}}$ belonging to the language $L(G)$ of the grammar $G$, along with the set of next valid tokens $M_P(y_{\text{prefix}})$. Then, a specific $\omega^*$ in $M_P(y_{\text{prefix}})$ is obtained via the language model and appended to $y_{\text{prefix}}$ to form a new DSL prefix, iteratively, until $\hat{y}$ is in $L(G)$, as illustrated in Figure 28.

*Constrained-Generation* $(x \in \Sigma^*, G : \text{CFG})$:

$\hat{y} \leftarrow \varepsilon$ ▷ Initialize empty string $\varepsilon$
**while True:**
    $\bar{y} \leftarrow \text{decode } P_{\text{LLM}}(y \mid x, G, \hat{y}, ...)$
    $\hat{y} \leftarrow \hat{y} \cdot \bar{y}$ ▷ Concatenate and update $\hat{y}$
    **if** $\hat{y} \in L(G)$**:** ▷ Try to verify the string $\hat{y} \in L(G)$
        **return** $\hat{y}$ ▷ Return expected DSL
    **else:**
        $y_{\text{prefix}}, M_P(y_{\text{prefix}}) \leftarrow \text{Generator}(\hat{y}, G)$
        $\omega^* \leftarrow \arg\max P_{\text{LLM}}(\omega \mid \omega \in M_P(y_{\text{prefix}}), y_{\text{prefix}}, ...)$
        $\hat{y} \leftarrow y_{\text{prefix}} \cdot \omega^*$ ▷ Update $\hat{y} \in P(G)$

Figure 28: Constraint generation (Wang, Wang, Wang, Cao, A Saurous, and Kim 2024)

Existing industrial practices and academic research have not yet adopted coroutines to implement the Generator module as described in Figure 28. It is easy to see that the algorithm mentioned requires validation of $\hat{y} \in L(G)$ in each autoregressive step. Even if the time complexity of validating $\hat{y}$, denoted by $n = |\hat{y}|$, is $O(n)$ (which is the best-case scenario and also the lower bound of complexity), the complexity of generating a string of length $\hat{y}$ is at least $O(n^2) = \sum_{i=1}^{n} O(n)$.

The coroutines utilized in this study naturally retain the context state of DSL parsing by allowing suspension and resumption during parsing or generation processes. This eliminates the need to scan the entire string from scratch for every validation or generation of new tokens, thus providing technical support for reducing the time complexity of constraint decoding.

## 5.4 Sampling with DSL Generator and Constraint Module

So how to specifically design a coroutine-based DSL generator and use it as a constraint module for sampling the domain-specific language as expected by developers using language models? This paper proposes YieldLang, a coroutine-based DSL generator implemented using Python metaprogramming techniques. The design principle of YieldLang is to uniformly process symbols of formal grammar into modules that can iteratively generate strings, while tracking the call stack of symbols for providing precise error and recovery information during DSL generation. Additionally, when YieldLang's generator is coupled with a language model sampler, its combinator functions can invoke large models to sample paths from a finite set of possibilities and recursively generate DSL through descent.

### 5.4.1 A Concrete Implementation of DSL Generator

This article implements a coroutine-based DSL generator in Python, with a key point being the recursive function `__flatten` within the `class TextGenerator`, as shown in Figure 29. It utilizes metaprogramming techniques to uniformly transform symbols (including terminals and non-terminals) Symbol from the formal grammar into iterable modules that generate strings, employing mechanisms such as delegation (yield from). Symbol can be a Strable that can be converted to a string, specific Tokens other than Unicode, empty (including None, empty strings, tokens representing emptiness, etc.), iterable that generates the aforementioned types, and Lambda functions capable of generating the aforementioned types.

```
def __flatten(self, symbol: Symbol) -> Iterable[str]:
    if callable(symbol):
        symbol = symbol()
    if symbol is None:
        yield EmptyString
    elif isinstance(symbol, Strable):
        yield str(symbol)
    elif isinstance(symbol, Token):
        yield from self.__process_token(symbol)
    elif isinstance(symbol, SamplerCallData):
        yield from self.__process_sampler(symbol)
    elif isinstance(symbol, Iterable):
        yield from self.__process_iterable(symbol)
    else:
        raise ValueError(f'Invalid symbol: {symbol}')
```

Figure 29: YieldLang unifies symbols into a module that can iteratively generate strings

If this module can only generate legal DSL, it seems a bit disappointing. Fortunately, `track_symbol` function is a decorator used to record symbol entry and exit situations, as shown by Figure 30. It records the function call parameters of recursive descent into a call stack. Outside the generator, it is always possible to obtain information such as symbol names from the top of the call stack. When the call stack changes, the generation status of DSL can be known, and of course, the entire call stack can be traced and restored to a syntax tree.

```
def track_symbol(fn: T) -> T:
    @functools.wraps(fn)
    def wrapper(*args, **kwargs):
        self = args[0] if args else None
        if isinstance(self, TextGenerator):
            self.sampler.call_stack.append((fn, args, kwargs))
            yield from fn(*args, **kwargs)
            self.sampler.call_stack.pop()
        else:
            fn_name = fn.__name__
            class_name = TextGenerator.__name__
            raise ValueError(f'{fn_name} is not a method of {class_name}')

    return wrapper
```

Figure 30: YieldLang tracks symbols in and out of decorator functions

In addition, the `class TextGenerator` of this apparatus also uniformly defines the Top function interface, representing the starting character $S$ in a formal language, which serves as the unique entry point for generating DSLs. Through Python's metaprogramming techniques, such as decorators, combinators, functors, and other methods, a framework for developing applications to generate DSLs can be implemented. Based on this framework, developers can write metalanguages on top of Python syntax. This framework is referred to as YieldLang.

### 5.4.2 Some DSL Generators for YieldLang

```
class PairsGenerator(TextGenerator):
    def top(self):
        yield self.pairs

    def pairs(self):
        yield select(
            (self.pair),
            (self.pair, self.pairs)
        )

    def pair(self):
        yield optional(
            '(', self.pairs, ')'
        )
```

Figure 31: A legal bracket pair generator for YieldLang

As shown by Figure 31, this is a generator for producing valid pairs of brackets. Its `top` function serves as the entry point of the generator, which calls the `pairs` function, and the `pairs` function in turn calls the `pair` function. The `pair` function is responsible for generating two types of possible bracket pairs: an empty string and a pair of brackets. This generator can produce any number of valid bracket pairs, such as `()`, `(())`, `(())()`, `(()())()`, and so forth. The `select` combinator function is used to select a sampling path, and the `optional` function is equivalent to the code shown by Figure 32.

```
def optional(self, *args: Symbol):
    yield select(
        (''),
        args
    )
```

Figure 32: The definition of optional combinator in YieldLang

Another more complex example is generating simple Mermaid flowcharts, the generator code of which is shown in Figure 33. In it, the Python expression `(yield self.graph_name)` will yield a symbol-generated string, and then based on the content of this string, it selects to generate `flowchart` (as a simple example, no other types of Mermaid charts are generated here). It is worth mentioning that `(yield A)` is equivalent to `yield (yield A)`. The `join` function is used to insert a specific symbol between each adjacent pair of symbols in the outermost symbol sequence. The `repeat` function is used to repeatedly generate a symbol a specified number of times.

```python
class MermaidGenerator(TextGenerator):
    def top(self):
        yield self.mermaid

    def mermaid(self):
        match (yield self.graph_name):
            case 'flowchart':
                yield self.flowchart

    def graph_name(self):
        yield select('flowchart')

    def flowchart(self):
        yield (' ', self.flowchart_type, '\n')
        yield join('\n', self.flowchart_rules)

    def flowchart_type(self):
        yield select('TD', 'LR')

    def flowchart_rules(self):
        rand_times = randint(10, 20)
        single_line = (' '*4, self.flowchart_rule)
        yield from repeat(single_line, rand_times)

    def flowchart_rule(self):
        yield self.node
        yield ' --> '
        yield self.node

    def node(self):
        yield select(*range(1, 10))
```

Figure 33: A simple flowchart generator for YieldLang

If a random sampler such as `class RandomSampler(Sampler)` is employed (which selects a random path at each intersection recursively in the descent sampling), then the generator based on YieldLang for Figure 33 can produce flowcharts with random nodes and random directed edges as illustrated in Figure 34.

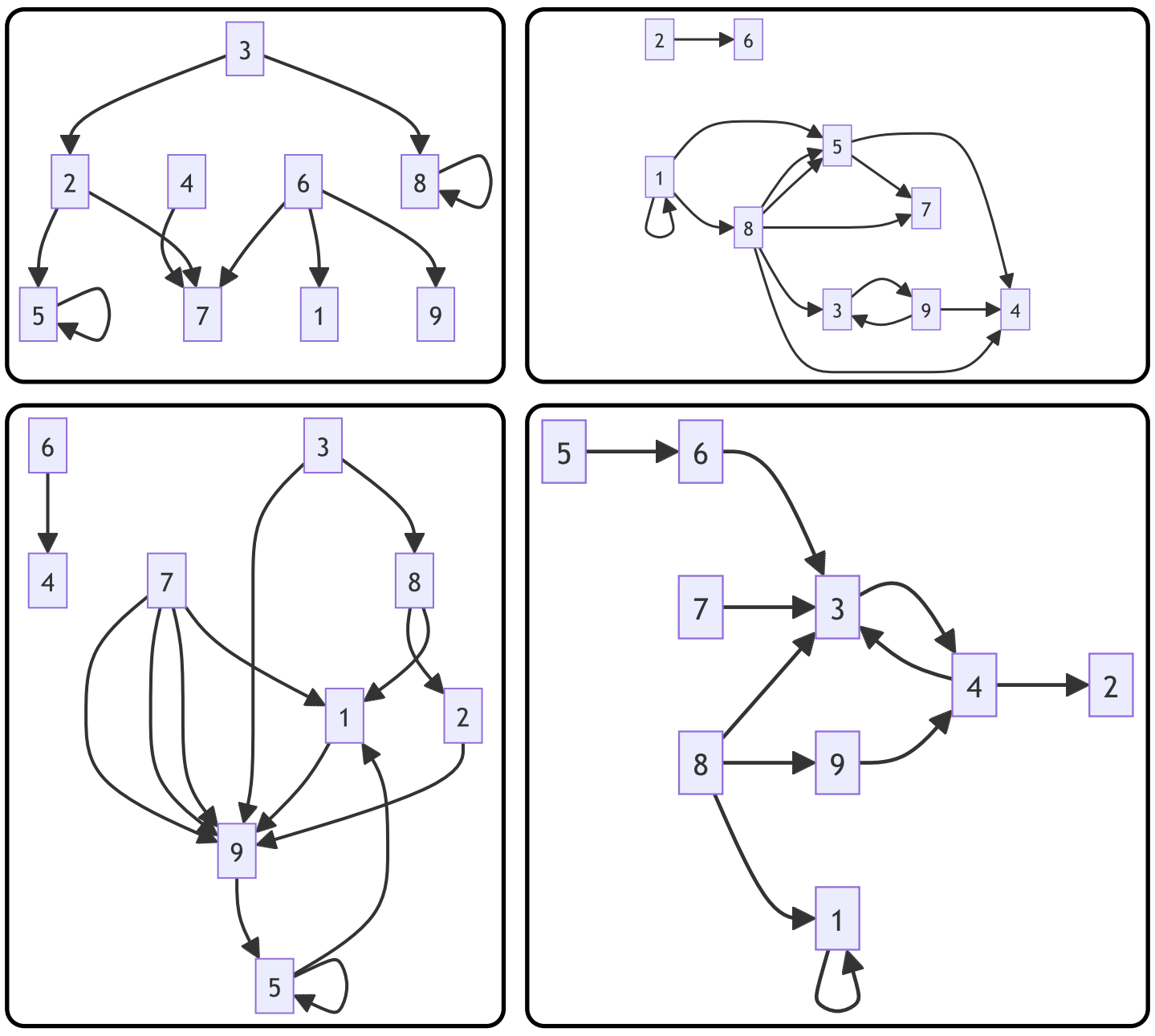

Figure 34: Four flowcharts randomly generated by Mermaid Generator

As demonstrated by the `print_sample_rules` function shown by Figure 35, random sampling of production rules from the CFG of this generator can be achieved through random sampling and the `track_symbol` function. This method can be employed to generate syntax trees for DSLs or production rule diagrams (rendered using Mermaid). Figure 27 is an example of a JSON syntax production rule diagram generated using this method.

```python
def print_sample_rules():
    rules: set[str] = set()
    for _ in range(sample_times):
        sampler = RandomSampler()
        for _ in JSONGenerator(sampler):
            last_symbol = None
            for symbol_name in sampler.symbol_names():
                if last_symbol and symbol_name:
                    rule = f'{last_symbol} --> {symbol_name}'
                    if rule not in rules:
                        rules.add(rule)
                        print(rule)
                last_symbol = symbol_name
```

Figure 35: Randomly sample rules from existing generators

JSON, being a widely used DSL in industry, can evidently benefit from the syntax definition and corresponding generator apparatus provided by YieldLang proposed in this paper. Figure 36 provides the complete generator code for JSON syntax, which is written according to the official McKeeman Form for JSON.

The `accept` function in JSONGenerator is used to define the character set that this symbol can accept (or the terminal symbol set that it can generate accept$(...) \subseteq \Sigma$). The `invalids` parameter is used to define the set of characters that this symbol cannot accept, while the `range` parameter is used to define the character range that this symbol can accept. With JSONGenerator in place, random sampling can generate random JSON strings of any length.

| JSON Samples | Object Types | | String Length |
|---|---|---|---|
| `null` | `none` | (null) | 4 |
| `"⎍"` | `string` | (string) | 5 |
| `false` | `boolean` | (boolean) | 5 |
| `-39.01` | `float` | (number) | 6 |
| `-96008` | `integer` | (number) | 6 |
| `"ꙍ責"` | `string` | (string) | 8 |
| `"/鷀?\""` | `string` | (string) | 9 |
| `[[[true]]]` | `array` | (array) | 10 |
| `{"꾋": []}` | `dictionary` | (object) | 13 |

Table 7: Sample JSON string generated by random sampling

Table 7 provided various examples of randomly sampled JSON strings, each with different types and lengths. It is important to note that these random JSON string examples lack any semantic meaning; they solely represent their syntactic structure, where the "garbage characters" result from randomly selecting characters from the Unicode character set.

```
class JSONGenerator(TextGenerator):
    def top(self):
        yield self.json
    def json(self):
        yield self.element
    def object(self):
        yield select(
            ('{', self.ws, '}'),
            ('{', self.members, '}')
        )
    def members(self):
        yield select(
            (self.member),
            (self.member, ',', self.members)
        )
    def member(self):
        yield (self.ws, self.string, self.ws, ':', self.element)
    def array(self):
        yield select(
            ('[', self.ws, ']'),
            ('[', self.elements, ']')
        )
    def elements(self):
        yield select(
            (self.element),
            (self.element, ',',  self.elements)
        )
    def string(self):
        yield ('"', self.characters, '"')
    def characters(self):
        yield optional(self.character, self.characters)
    def character(self):
        yield select(
            accept(range=('\u0020', '\uffff'), invalids=('"', '\\')),
            ('\\', self.escape)
        )
    def escape(self):
        yield select(
            *'\\"/bfnrt',
            ('u', repeat(self.hex, 4))
        )
    def hex(self):
        yield select(
            self.digit,
            select(*'ABCDEF'),
            select(*'abcdef')
        )
    def digit(self):
        yield select('0', self.onenine)
    def onenine(self):
        yield select(*'123456789')
    def number(self):
        yield (self.integer, self.fraction, self.exponent)
    def integer(self):
        yield select(
            (self.digit),
            (self.onenine, self.digits),
            ('-', self.digit),
            ('-', self.onenine, self.digits)
        )
    def digits(self):
        yield select(
            (self.digit),
            (self.digit, self.digits)
        )
    def fraction(self):
        yield optional('.', self.digits)
    def exponent(self):
        yield optional(select(
            ('E', self.sign, self.digits),
            ('e', self.sign, self.digits)
        ))
    def sign(self):
        yield optional(select('+', '-'))
    def boolean(self):
        yield select('true', 'false')
    def null(self):
        yield 'null'
    def value(self):
        yield select(
            self.object,
            self.array,
            self.string,
            self.number,
            self.boolean,
            self.null
        )
    def element(self):
        yield (self.ws, self.value, self.ws)
    def ws(self):
        yield optional(select(
            ('\u0020', self.ws),
            ('\u000A', self.ws),
            ('\u000D', self.ws),
            ('\u0009', self.ws)
        ))
```

Figure 36: A JSON text generator for YieldLang

### 5.4.3 How to Guide Language Models in Sampling Paths

Having just a random sampler is certainly insufficient. To leverage the capabilities of language models to generate truly semantic Domain Specific Languages, it's necessary for the language model to choose sampling paths. However, while the neural network output layer of the language model can represent the probability distribution $p(x_{n+1})$ of the next word in the vocabulary $\Sigma$, it cannot directly provide a probability distribution for non-terminals. This means that we cannot directly instruct the language model to sample paths according to rules; instead, we first need to find the "forks" where each fork corresponds to a non-terminal $A \in N$ and determine the First set for each non-terminal, which dictates the first symbols it can produce (if a path directly leads to a terminal symbol, it can be selected directly).

In formal languages and compiler theory, the concept of First sets is crucial for predicting the next input symbol of a non-terminal in a context-free grammar. The First set contains the set of symbols that a non-terminal can potentially produce as the first symbol and is used to predict whether a non-terminal can derive a certain symbol.

For a non-terminal $A$, its First set, denoted as $\text{First}(A)$, is defined by the following three rules:

1. If $A$ can directly derive the empty string $\varepsilon$, then $\varepsilon \in \text{First}(A)$.
2. If $A$ can derive a string starting with a terminal symbol $a \in \Sigma$, then $a \in \text{First}(A)$.
3. If $A$ can derive a string starting with a non-terminal $B$, then $\text{First}(B) \subseteq \text{First}(A)$.

It's apparent that based on the three rules of First sets, an iterative algorithm can be designed to compute the First set. The basic idea of this algorithm is to initialize $\text{First}(A)$ as the empty set and then iterate until $\text{First}(A)$ no longer increases. In each iteration, for each production $A \to \alpha$ in $P$, if $\alpha$ is a terminal or an empty string, then $\alpha$ is added to $\text{First}(A)$. If $\alpha$ is a non-terminal, then all elements of $\text{First}(\alpha)$ are added to $\text{First}(A)$. The time complexity of this algorithm is $O(|N| * |P|)$, where $|N|$ is the number of non-terminals and $|P|$ is the number of productions. For the context-free grammar defined by the quadruple $G = (N, \Sigma, P, S)$, the First sets can be computed using the basic algorithm shown in Figure 37.

```
First (A ∈ N):
s ← ∅
for (A → α) in P:
    if (α → ε):
        s ← s ∪ {ε}
    else if (α = a ∈ Σ):
        s ← s ∪ {a}
    else if (α = B ∈ N):
        s ← s ∪ First(B)
return s
```

Figure 37: An algorithm for finding the First set

The problem with the algorithm demonstrated by Figure 37 is that it cannot handle cases of left recursion. Left recursion occurs when a non-terminal starts with itself in the right-hand side of a production, for example, $\{A \to Aa, A \to b\}$. In such cases, computing the First set for $A$ will lead to an infinite loop. This is because the First set of $A$ may contain $A$ itself, so computing the First set of $A$ requires computing the First set of $A$ again, leading to an endless loop. Additionally, if multiple symbols form an indirect left recursion, a similar situation of endless looping will occur.

To eliminate the problem of left recursion, there are two methods: one is for developers to avoid left recursion when writing the grammar of the DSL, and the other is to automatically eliminate left recursion during the construction of the generator based on the grammar rules. Of course, for LLMs with general capabilities, it is acceptable if the First set cannot be computed. Instructions can be added to allow the language model to directly select non-terminals (since non-terminals themselves can be converted into strings, and developers can even describe these non-terminals, thus constructing one or more strings belonging to the alphabet $\Sigma$ of the LLM for sampling). This enables the language model to sample paths from the DSL syntax.

The system demonstrated by Figure 3 for guiding LLMs to generate parseable content also utilizes the language model as a sampler for DSL generation, as shown in the process diagram. The process is as follows: i. Input prompts, and the language model begins generating from an empty context. ii. The asynchronous DSL parsing and language generation module produce guiding instructions, which are transformed by the language model guidance module into the next set of legal tokens. iii. The language model guidance module acts on the output layer of the NN (neural network), and the Tensor of the neural network output layer is transformed into a new Tensor by a post-processor, then converted into a probability representation of the language model vocabulary through the softmax function, and finally a specific token is obtained through sampling. iv. The language model guidance module utilizes the Tokenizer corresponding to the language model's vocabulary to obtain the next set of legal tokens. v. Vocabulary bias: The language model guidance module inserts a new constraint processor between Logits and New Logits to increase the values of legal ID positions in the probability distribution Tensor and decrease the values of illegal ID positions. vi. Through the decoder of the model, the model outputs the correct new token, forming a new DSL or prefix of the DSL. vii. Input the token to the asynchronous module through the send method, and obtain new guiding instructions. viii. Return to step iii, continue generating based on the context of the current task, until the asynchronous module generates an end instruction, indicating that the expected DSL generation by the developer is complete and this DSL can be parsed.

## 5.5 Summary of This Chapter

This chapter first introduces the autoregressive models commonly used by language models, and then illustrates the generation approach for JSON strings using JSON syntax as an example, pointing out the shortcomings of current common parsers. Subsequently, mathematical definitions are provided for prefix language $P(G)$ and module $M_P(s)$. Following this, the chapter introduces the constrained decoding strategy for autoregressive language models. The text then describes a DSL generation framework designed in Python, named YieldLang, and provides examples of CFGs written in YieldLang as metalanguages. Finally, the chapter outlines a method for language models to select sampling paths in the generator and the First set, ultimately providing a detailed procedural description of a system guiding LLMs to generate computer-parsable content.

# 6 Conclusions

We tested the abilities of the GPT-2 and Gemma models in downstream tasks of DSLs on three test sets, with the temperature set to 0.7. The JSON Text task considers the accuracy of the models in generating specific JSON subsets under the complete JSON grammar. The Mermaid task evaluates the success rate of the models in generating vector graphics that can be rendered into SVG format by Mermaid.js from a subset DSL of flowcharts. The Function Call task assesses the success rate of the models in generating Python function call expressions that can be successfully invoked.

The performance of the latter significantly outperforms the former in both scenarios: without the constraints of DSL (Origin) and with the use of YieldLang implemented in this paper (Ours), as demonstrated by Table 8.

| Model Names | JSON Text | | Mermaid | | Function Call | |
|---|---|---|---|---|---|---|
| | Origin | Ours | Origin | Ours | Origin | Ours |
| GPT-2 | 6.7% | **12.1%** | 7.2% | **83.6%** | 16.7% | **18.9%** |
| GPT-2 XL | 13.5% | **19.6%** | 11.3% | **87.4%** | 19.0% | **20.7%** |
| Gemma-2B | 20.4% | **42.1%** | 23.2% | **91.1%** | 23.7% | **26.4%** |
| Gemma-7B | 29.3% | **49.9%** | 34.4% | **97.7%** | 28.2% | **31.9%** |

Table 8: The scores of this study on various DSL generation tasks

Since certain paths in the generator's production of DSLs can always be determined by the program without the need for extensive sampling by the LLM, this saves the sampling iterations required for DSL generation by the LLM. Experimental results indicate that in the scenario of generating semantically inert JSON strings, under the most favorable conditions (when $\Sigma_{\text{LLM}} \subseteq$ Unicode), the sampling iterations can be reduced to approximately 16.5% of the original amount, as shown by Figure 38, varying with the string length.

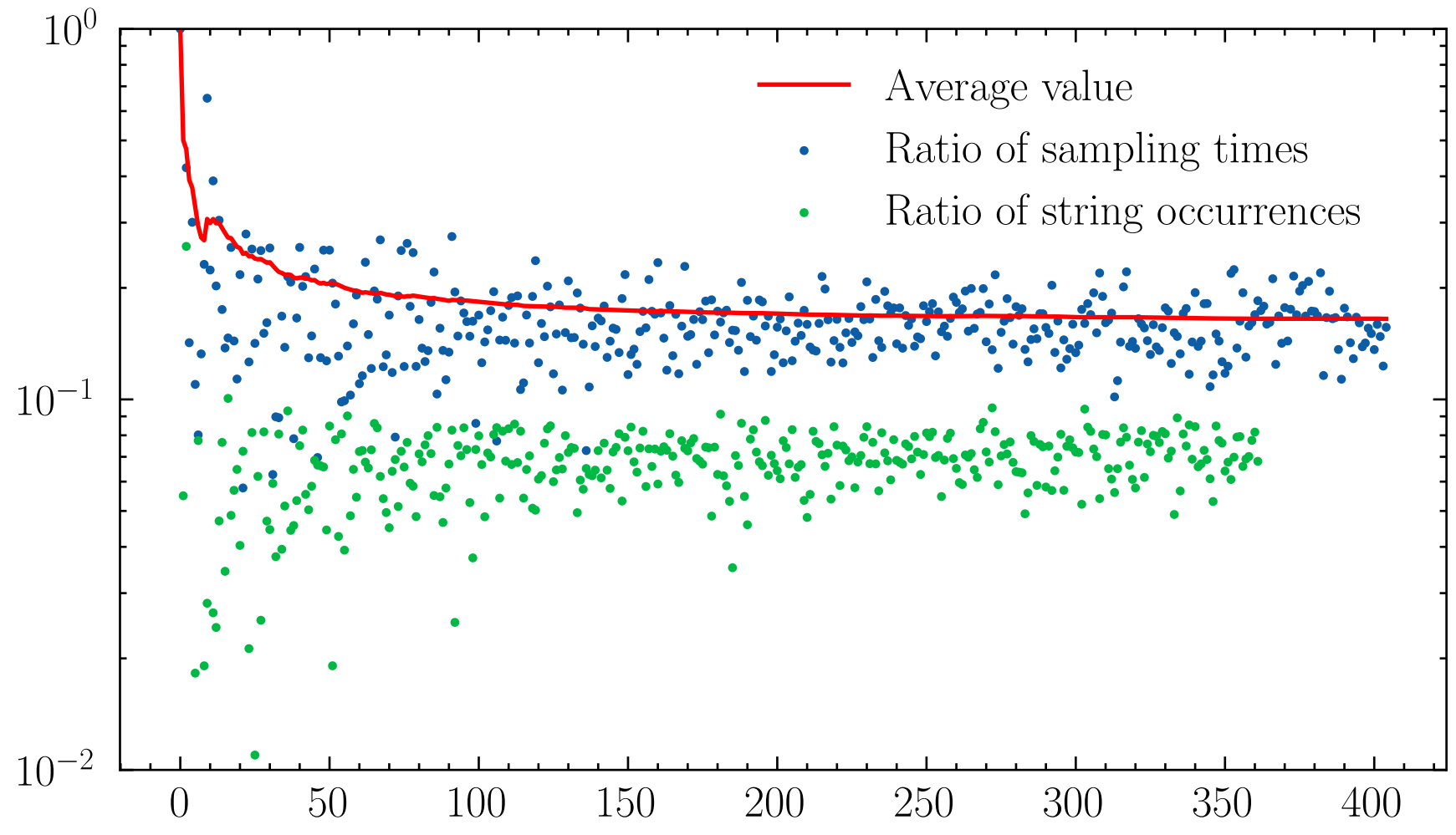

Figure 38: The ratio of the number of samples to the length of the generated JSON

If only considering DSLs that can be successfully parsed by a parser, this study conducted around 270 million tests on JSON generation and parsing using Python's JSON package. It generated JSON strings with an average length of 17.86 and a maximum length of 2136, where 100% of the JSON strings were successfully parsed and restored.

## 6.1 Implementing an LLM's JSON Mode

JSON, as a typical Domain-Specific Language, is capable of representing common data structures. In this study, based on YieldLang and the syntax rules of JSON, we implemented the JSON mode for LLMs, as demonstrated in Figure 39.

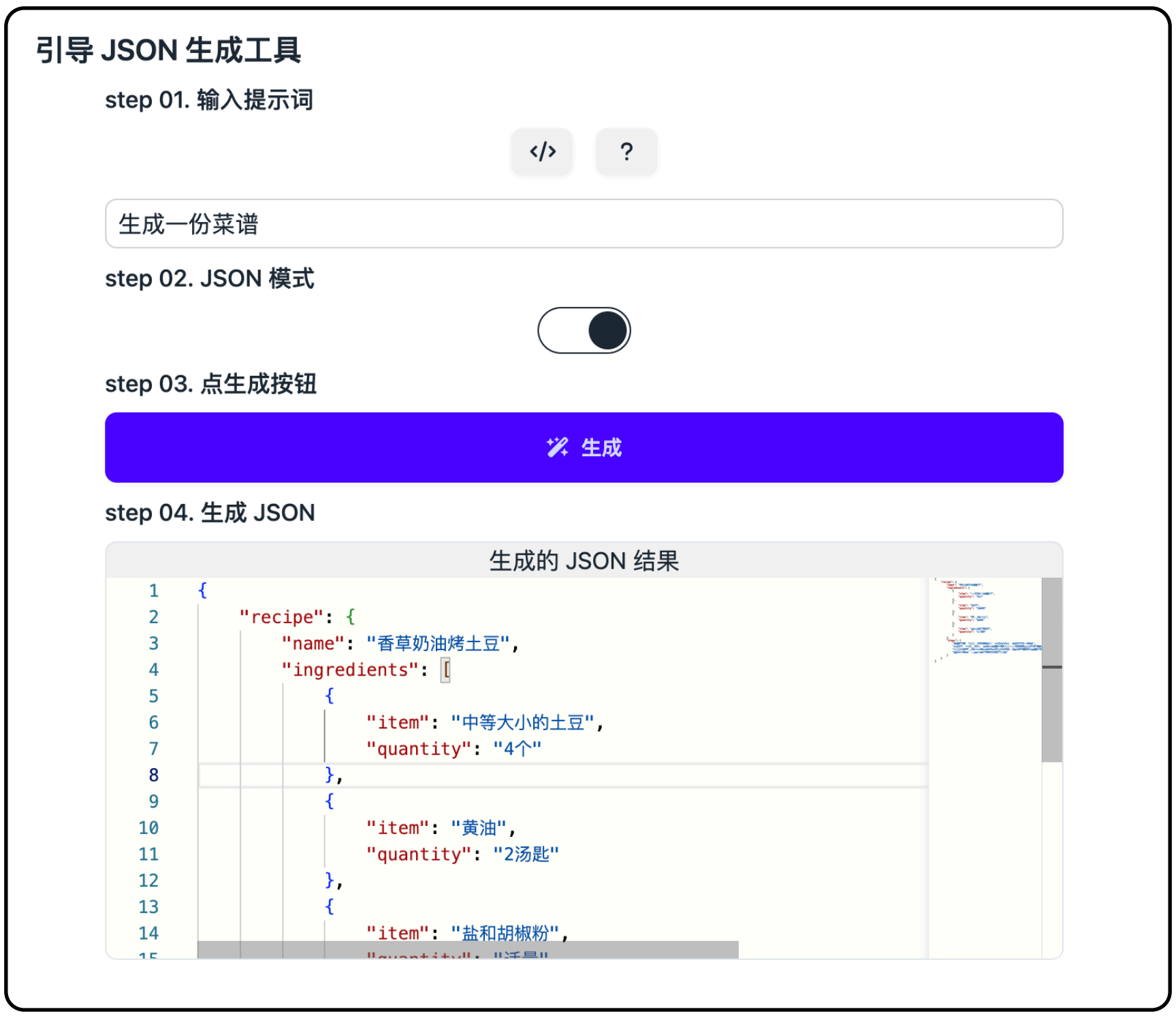

Figure 39: A Simple UI for LLM's JSON mode

The user interface of Figure 39 consists of several components from top to bottom: a title, a button for entering prompt words, an input box for prompt words, a toggle switch for enabling JSON mode, a generation button, and a preview area for the generated result. The first button for prompt word input is used to edit the Markdown source code of prompt words, and the second button is for accessing help on using the input box. The input box is used to enter prompt words provided to the language model. Disabling JSON mode means that the model will not utilize the apparatus proposed in this study. Enabling JSON mode activates the proposed apparatus and configures the Context-Free Grammar (CFG) for JSON. Clicking the generation button prompts the language model to sequentially generate tokens based on the prompt words and append them to the end of the cursor in the result preview editor until the model outputs an end token, indicating the completion of JSON generation by the LLM.

In industrial practice, the apparatus proposed in this study can be incorporated into Transformer-based inference frameworks, providing an API similar to that of OpenAI, for example, `{"response_format": {"type": "json_object"}}` indicating that the language model should return the response in the format of a `json_object`. Furthermore, techniques such as prompt engineering can be employed to activate the DSL constraint apparatus at "appropriate" times or "positions" to achieve specific functionalities.

# 7 Summary and Prospects

This study examines the challenges and methods of generating domain-specific languages (DSLs) using LLMs. Firstly, it introduces the widespread application and current research status of DSLs, highlighting the potential and limitations of language models in generating DSLs. The typical syntactic features of DSLs are analyzed, and an experimental method for evaluating the performance of language models in DSL generation is proposed. Experimental results demonstrate that existing language models struggle to achieve acceptable performance on DSL generation tasks. To address this issue, a constraint decoding-based DSL generation module is proposed, consisting of an asynchronous DSL parsing language generation module and a language model-guided module. This study also introduces a constraint decoding strategy for autoregressive language models and designs the YieldLang framework, a user-friendly DSL generation framework based on Python's metaprogramming capabilities.

The devised approach in this study demonstrates advantages in multiple DSL generation tasks, enhancing the performance of LLMs in generating DSLs. Remarkably, the devised module in this study is expected to reduce the sampling frequency of LLMs, thereby reducing the computational burden and time required for DSL generation by language models. Additionally, a JSON schema for LLMs has been implemented, providing a human-computer interface and a web API where users can generate specific JSON strings by providing prompts to the LLM. The recent emergence of MoonBit language has begun to explore the use of DSL samplers to "adapt" LLMs[12], and the coroutine-based approach proposed in this study holds promise to enhance existing language model DSL generation modules and reduce development complexity.

In essence, this study can be roughly considered as a method of using "autocompletion" to assist or constrain language models in generating DSLs. Constraint decoding can be metaphorically understood as "speaking to the language model with constraints". Furthermore, DSLs, as a unified, general, and widely applied form, hold the potential to help artificial intelligence grasp the usage of tools and become AGI. Two potential application scenarios are listed below.

1. Autocompleter: Using language models to assist developers in implementing requirements with fewer steps. If the language model can construct more specific DSL syntax rules based on contextual environments and then assist developers in writing code through constraint decoding, it may significantly increase developer adoption rates.
2. Intelligent operator: Generating 100% correct function call expressions or other DSLs based on CFGs, allowing the model to call functions with a higher success rate to accomplish specific tasks based on contextual environments.

Finally, there are numerous shortcomings in this study, such as how LLMs perform under more complex constraints and what their performance and conditional probabilities are, how to integrate prompt engineering techniques, and In-Context Learning methods with DSL constraints to enable the model to perform well in more scenarios, and whether related methods are beneficial for LLMs' pre-training or fine-tuning. These questions will be further explored in future research.

[12] https://www.moonbitlang.com/blog/moonbit-ai